\documentclass[english,a4paper,11pt]{article}
\usepackage[margin=3cm]{geometry}
\usepackage{babel}
\usepackage{bm}
\usepackage{latexsym}
\usepackage{float}
\usepackage{booktabs}   
\usepackage{multirow}   
\usepackage{array}  
\usepackage{amsmath,amsthm}
\usepackage{latexsym}
\usepackage{booktabs}
\usepackage[final]{graphicx}
\DeclareGraphicsExtensions{.jpg,.jpeg,.pdf,.png,.mps}
\usepackage{epsfig}
\usepackage[round]{natbib}
\usepackage{rotating}
\usepackage{color}
\usepackage{xcolor}
\usepackage{blindtext}
\usepackage[colorlinks=true, linkcolor=blue, citecolor=blue, urlcolor=blue]{hyperref}
\usepackage{amsfonts}
\usepackage{colortbl}
\usepackage{hyperref}
\usepackage[misc]{ifsym}
\usepackage[T1]{fontenc}
\usepackage{lmodern}

\title{The effect of a new power interconnector on energy prices volatility: the case of Sicily}

\author{$\mathrm{Francesco \ Lisi}^\mathrm{1,2}, \ \mathrm{Pierdomenico \ Duttilo}^\mathrm{1,\hspace{0.5mm}\textrm{\Letter}},  
	\   \mathrm{Marina \ Bertolini}^\mathrm{1,2}$
	\\  
    $^\mathrm{1}$\small{\emph{Department  of Statistical Sciences,  University of Padua, Italy}}\\
    $^\mathrm{2}$\small{\emph{Interdepartmental Centre ``Giorgio Levi Cases'' for Energy Economics and Technology,}}\\
     $^\mathrm{}$\small{\emph{University of Padua, Italy}}\\
    $^\mathrm{\textrm{\Letter}}$\small{{Corresponding author: \href{mailto:pierdomenico.duttilo@unipd.it}{\textcolor{blue}{pierdomenico.duttilo@unipd.it}}}}
}

\date{}

\begin{document}
	\maketitle

\begin{abstract}
	\noindent Integrating energy islands into the European electricity market is a key challenge for the energy transition. This study investigates the impact of the ``Sorgente-Rizziconi'' interconnector on electricity price volatility in Sicily. Before its commissioning on 28 May 2016, the Sicilian electricity market zone was poorly interconnected with the Italian mainland. Using daily data from 2015 to 2018, the analysis applies a semi-parametric GARCH model with a logistic intervention function to estimate changes in conditional price variance. A fully non-parametric additive model is employed as a robustness check, allowing the data to shape volatility dynamics without imposing a predefined structure. The results reveal that the new interconnector significantly increased price volatility in Sicily, without reducing average price levels. No significant effects were observed in other Italian market zones. These findings highlight the context-dependent nature of infrastructure impacts and suggest that physical integration alone does not guarantee price stability. The results have important implications for energy policy, investment planning, and risk management in electricity markets.\\
	\hspace{1cm}
\end{abstract}

\section{Introduction}
\label{sec:1}
The integration of European electricity markets represents a central target of the European Union since the first Energy Package adopted in 1996 \citep{roques2021integration}. In October 2014, the European Council approved the 2030 Climate and Energy Framework to promote the integration of the internal energy market through coordinated decisions on renewables, balancing markets, production capacity, and interconnector development \citep{EC2014climateframework}. Furthermore, Directive (EU) 2024/1711 reaffirms the commitment \citep{EU2018Reg1999_governance,EU2024directive1711} of the European Union to ensure that each member state achieves an electricity interconnectivity level of 15\% by 2030. A fully integrated energy market within the European Union offers important benefits and represents the most efficient approach to provide secure, sustainable, and affordable energy to all EU citizens \citep{CRETI20106966,NEWBERY2016253,CASSETTA2022112934}. Current levels of EU market integration are estimated to produce annual consumer savings of approximately €34 billion, with the potential to increase by 2030 \citep{EU_ElectricityMarketDesign}.

However, one of the main constraints to achieve a unified electricity market is the persistence of the so-called ``energy islands''. They can be classified in: physical, political, and service islands \citep{RETTIG2023113732}.\\ Physical energy islands are geographically remote or isolated areas with limited or no access to external electricity networks, such as Cyprus \citep{Proedrou201215}, Sardinia \citep{Sapio2020}, Sicily \citep{Sapio16} and Iceland \citep{SHORTALL2017101}.\\ Political energy islands refer to mainland countries or regions that remain electrically unconnected to neighbouring states despite their geographical proximity. This condition arises from geopolitical tensions, regulatory barriers, or insufficient infrastructure; examples include Israel and South Korea \citep{Fischhendler2015}.\\ Service energy islands encompass specific areas such as self-sufficient cities, regions, local communities, or artificial islands that are designed to operate independently or semi-independently from national or international grids \citep{WARNERYD2022,JANSEN2022}.\\ Each form of ``energy islanding'' presents different challenges and opportunities that highlight the different context in which energy isolation occurs and must be addressed to achieve greater market integration. Their integration requires investments in the development of interconnectors, i.e., ``an asset in the form of an underground or overhead transmission line whose purpose is to link and allow the transfer of electricity between two individual electric power systems'' \citep{RIES20161}. From a historical perspective, the world’s first underwater electrical transmission cable was installed in 1811 in the Isar River area of Bavaria, Germany. Since then, electricity interconnectors have undergone substantial evolution, driven by technological progress, improved design, and increased electricity demand \citep{ESCA_SubmarineCables}. Interconnectors play a crucial role in reducing the isolation of remote regions, enabling cross-border electricity exchanges, improving system reliability, and promoting price convergence within integrated electricity markets \citep{TURVEY20061457,CARTEA201214}.

\subsection{Background literature}
\label{sec:literature}
The installation of new power interconnectors represents a significant structural transformation within electricity systems. As a largely ``physical'' market, any modification to the energy infrastructure directly affects the operation of the system and the formation of equilibrium prices \citep{CHENG2025110588}. As a consequence, the configuration of the electrical grid influences the ability of operators to participate effectively in the market \citep{TSAOUSOGLOU2022}. Electricity price volatility can serve as an indicator of both market competitiveness \citep{Bertolini20} and the capacity of the network to integrate local producers \citep{TISHLER20081625}. It reflects the responsiveness of prices to changes in supply and demand, with higher volatility suggesting that the market can adjust more efficiently \citep{LI2025101808,MUGALOGLU2026}. Furthermore, price volatility can be interpreted as evidence of infrastructure enhancements, so called smart upgrades, designed to accommodate a larger number of market participants. This is particularly relevant for operators with intermittent or small-scale generation, whose integration requires additional investments by local grid operators to maintain system reliability and efficiency \citep{MILSTEIN201570,WANG2024144343}. Although higher price volatility may discourage risk-averse participants, it can also have a pro-competitive effect by increasing expected profits \citep{Bertolini20}.

However, both the direction and the magnitude of the effect of new interconnectors on electricity price volatility depend on the specific market context, and there are no general conclusions.

Interconnectors link previously isolated or weakly connected systems and promote price convergence and efficiency improvements \citep{TURVEY20061457, CARTEA201214}. By improving electricity exchanges between regions, interconnectors mitigate local supply shocks and reduce price disparities. This stabilising effect has been observed in the Irish and British markets, where increased interconnection improved competition and contributed to a reduction in both prices and their variability \citep{MALAGUZZIVALERI20094679, DENNY20106946}. Further evidence from studies on the Franco-British interconnector highlights the role of these infrastructures in enhancing operational flexibility, reducing renewable curtailment, and supporting a more stable price formation process \citep{PEAN2016307}. Similar benefits have been identified in the context of Spanish islands, where interconnectors have improved the reliability and cost-efficiency of the system, leading to a smoother price dynamics \citep{LOBATO2017192}.

Although new interconnectors contribute to market efficiency, they can also transmit volatility across regions, particularly when connected systems differ in generation structure or penetration of renewables \citep{eu_grid_action_plan_2023}. Evidence from the Italian market shows that the commissioning of the SAPEI cable between Sardinia and mainland Italy intensified volatility spillovers, especially during off-peak periods, despite improved average price convergence \citep{Sapio2020}. Specifically, volatility tended to flow asymmetrically from larger zones with high shares of intermittent renewables (particularly the Central Southern zone) towards Sardinia, a smaller net-importing market. In contrast, during peak periods, congestion on the SAPEI line constrained volatility spillovers. The underlying mechanism reflects the physical reality of electricity markets, where the limits of transmission capacity and uneven system conditions can convert integration into a source of instability \citep{TSAOUSOGLOU2022, FABRA2023, YANG2024}.

The absence of a clear impact can occur when existing market mechanisms, such as efficient dispatching or adequate balancing capacity, already ensure price stability. \citep{RIES20161}, analysing the Malta--Sicily interconnector, show that the benefits of interconnection depend heavily on the regulatory design and the mix of generation, with limited effects on consumer prices in certain scenarios. Similarly, \citep{DENNY20106946} report that although greater interconnection improved adequacy and reduced prices, it did not substantially alter other system dynamics, suggesting partial neutrality. This interpretation is consistent with the broader view that infrastructure benefits materialise only when supported by coherent regulatory and institutional frameworks \citep{LOPRETE2025108640, DUDJAK2021117434}.

\subsection{Study contribution}
This study investigates the effects of the Sorgente-Rizziconi interconnector on the volatility of the Sicilian electricity market. Sicily's electricity market was poorly connected with mainland Italy before 28 May 2016. 

\cite{Sapio16} analysed the pre-installation situation and suggested that the high electricity prices in Sicily were mainly driven by tacit collusion between market players. The specific effects of the Sorgente-Rizziconi interconnector on the market have not yet been explored. Since the literature offers divergent conclusions, this study aims to test the following hypotheses regarding the impact of the Sorgente-Rizziconi interconnector on electricity price volatility:
\begin{itemize}
\item Hypothesis 1 (\textbf{H1}): \textit{New interconnectors reduce electricity price volatility \citep{TURVEY20061457, CARTEA201214,MALAGUZZIVALERI20094679, DENNY20106946,PEAN2016307,LOBATO2017192};}
\item Hypothesis 2 (\textbf{H2}): \textit{New interconnectors increase electricity price volatility \citep{eu_grid_action_plan_2023,Sapio2020,TSAOUSOGLOU2022, FABRA2023, YANG2024};}
\item Hypothesis 3 (\textbf{H3}): \textit{New interconnectors do not have a significant effect on electricity price volatility \citep{RIES20161,DENNY20106946,LOPRETE2025108640, DUDJAK2021117434}.}
\end{itemize}

To address these hypotheses, while previous research on energy islands has mainly relied on deterministic simulation models \citep{MALAGUZZIVALERI20094679,DENNY20106946,CLEARY201638,PEAN2016307,RIES20161,LOBATO2017192}, this study contributes to the literature by combining stochastic modelling, empirical data analysis, and applications in electricity production. First, an exploratory data analysis is conducted to identify the most suitable models for electricity prices and volatility. Second, the conditional mean of electricity prices is estimated using a non-parametric additive model to filter out calendar and renewables-related effects. Third, two approaches are employed for the conditional variance: a GARCH-Logistic (GARCH-L) model for a parametric specification, and a non-parametric additive model as a robustness check for the GARCH-L results. The GARCH-L model is specifically designed to capture potential level shifts in conditional volatility, whereas the non-parametric additive model provides greater flexibility to account for possible non-logistic patterns. This combined approach enables the assessment of both the impact and duration of interconnector activation on price volatility, including whether it results in an increase or a decrease. The model hypothesises a sigmoidal adjustment pattern in which volatility begins to change at the time of activation and gradually stabilises at a defined plateau. Fourth, to assess the statistical significance of the effect of the interconnector on volatilty, two hypothesis tests are implemented: a t-test on the logistic coefficients of the GARCH-L model and an ANOVA test for the non-parametric additive model. Furthermore, results are compared across other Italian market zones for benchmarking purposes.

As expected, \textbf{H2} is verified in Sicily, while \textbf{H3} holds in the other market zones for the period 2015–2018. These findings are market zone-dependent with important implications for policy-makers, investors, and market players, as they provide information on how investments in energy-grid infrastructure influence electricity price volatility. In addition, they are relevant for business and industrial decision-making, since understanding the effects of interconnector activation can guide investment strategies, risk management, and operational planning in the power industry. These findings are market zone-dependent with important implications for policy-makers, investors, and market players.

The remainder of the paper is organised as follows. Section \ref{sec:setup} outlines the structure of the Italian electricity market. Section \ref{sec:data} describes the data. Section \ref{sec:models} details the modelling approaches. The results are presented in Section \ref{sec:results}. Finally, Section \ref{sec:conclusions} concludes.

\section{Italian market setup}\label{sec:setup}
\begin{figure}[H]
	\centering
	\includegraphics[width=15
	cm, height=8cm]{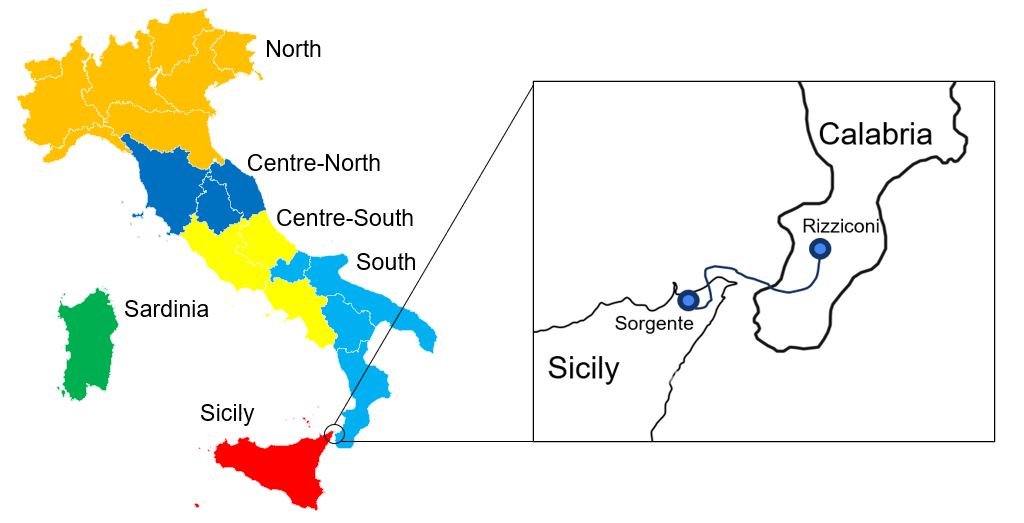}
	\caption{The six Italian electricity market zones with a detailed view on the Sorgente-Rizziconi interconnector.}
	\label{fig:zones_cable}
\end{figure}

The Italian power market is divided into zones, each representing a part of the national grid with physical limits on electricity transfers between zones. Terna\footnote{The Italian transmission system operator.} defines the market zones according to well-established principles. Until the end of 2020, before the latest revision\footnote{The specific zonal configuration is not relevant for this work.} implemented in 2021, the Italian electricity market was divided into six geographical zones: North, Centre-North, Centre-South, South, Sicily, and Sardinia. Figure \ref{fig:zones_cable} illustrates these six market zones, with a detailed view of the Sorgente–Rizziconi interconnector.

Between contiguous zones, transmission limits vary over time and define the maximum capacity for electricity transfers. These limits mainly result from constraints imposed by existing infrastructures. Under certain conditions, they can cause grid inefficiencies and bottleneck effects. Such constraints influence the dynamics of regional electricity markets, shaping market operations, and overall efficiency \citep{Weinhold21}. Within each zone, market exchanges occur and equilibrium prices are set for all stages of the market, including the day-ahead market. Network constraints affect all market mechanisms.

The design of a new interconnector linking Sicily to the mainland began in 2003. According to Terna\footnote{\url{https://download.terna.it/terna/0000/0125/44.pdf}}, the project was expected to generate savings in system management by increasing the exchange capacity between the South and Sicily. The connection aimed to better integrate renewable energy into the market, as a larger number of wind farms were connected to the Sicilian grid, allowing greater wind energy transfer to the mainland. The new interconnector was also expected to ease operational restrictions and enhance competition within the Sicilian electricity market zone.

Furthermore, the integration of renewable sources and improved network management were expected to reduce CO$_2$ emissions. The motivations behind the project were primarily technical, but also affected both management costs—linked to network operation and service quality—and energy costs, which depend on the degree of integration of renewable energy.

After more than a decade of work and an investment of approximately €700 million, the Sorgente–Rizziconi project was completed and opened on 28 May 2016. The interconnector crosses 23 municipalities in the provinces of Messina and Reggio Calabria. It is the longest alternating-current (380 kV) sub-submarine interconnector in the world. The total length is 105 km, including 38 km of submarine cables laid at depths of up to 376 metres. Its transmission capacity reaches 1,100 MW, sufficient to supply electricity to about three million homes.

Before industrial investment, a distinctive feature of the Sicilian electricity market was the high level of advance prices \citep{Sapio16}, which significantly influenced the Italian national price \citep{Meneguzzo2016}. Terna identified a reduction in local prices as one of the expected benefits of the new interconnector \citep{Terna2016}. However, as noted in contemporary news articles\footnote{See \cite{Codegoni2016} and \cite{Trovato2016}.} and confirmed by data analysis in this study, no clear price decrease was observed. It should also be noted that, in 2015, changes in merit order occurred due to the growing penetration of renewables in the region \citep{Meneguzzo2016}. Nevertheless, subsequent price trends did not reflect either the expansion of renewables or the expected effects of the new infrastructure.

\section{Data}\label{sec:data}
\begin{figure}[H]
	\centering
	\includegraphics[width=6.6cm, height=5cm]{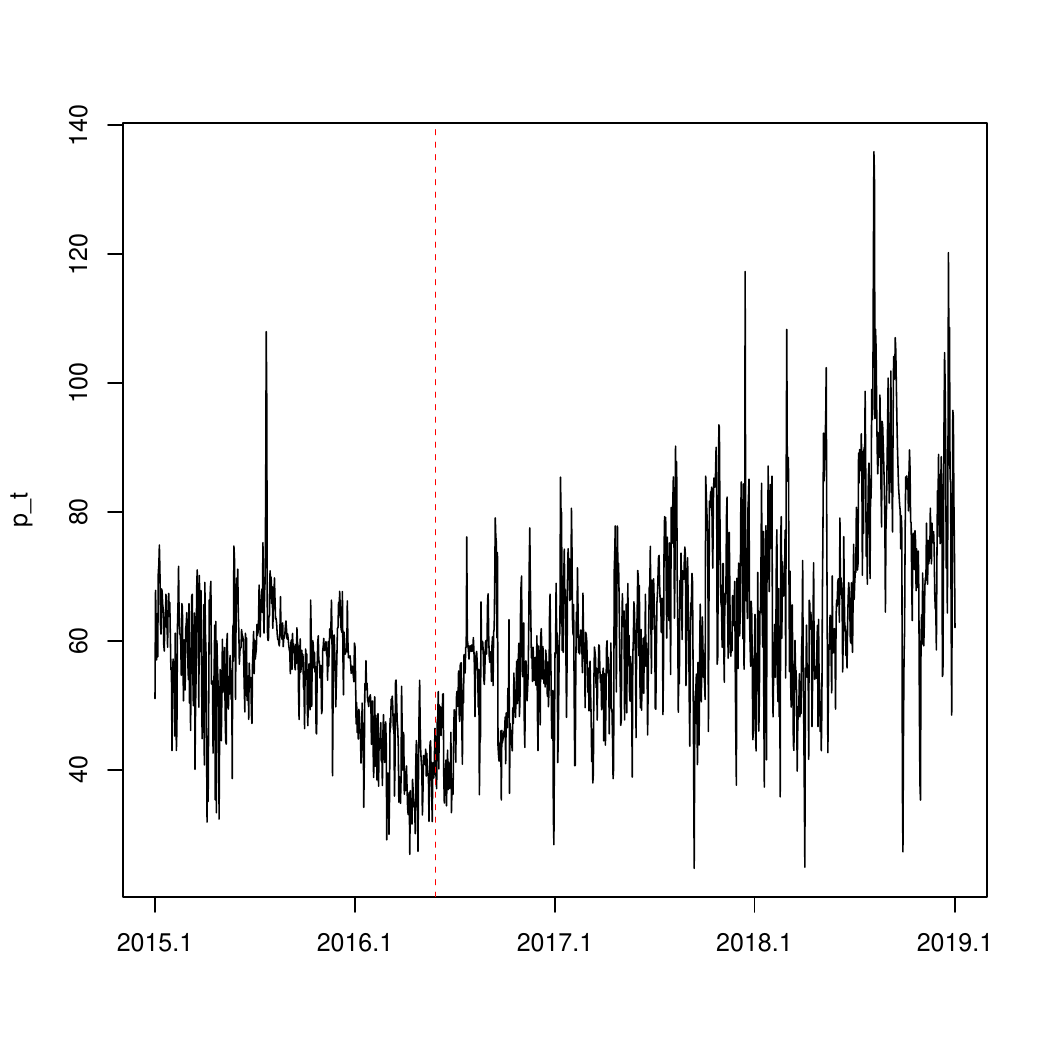}
	\includegraphics[width=6.6
	cm, height=5cm]{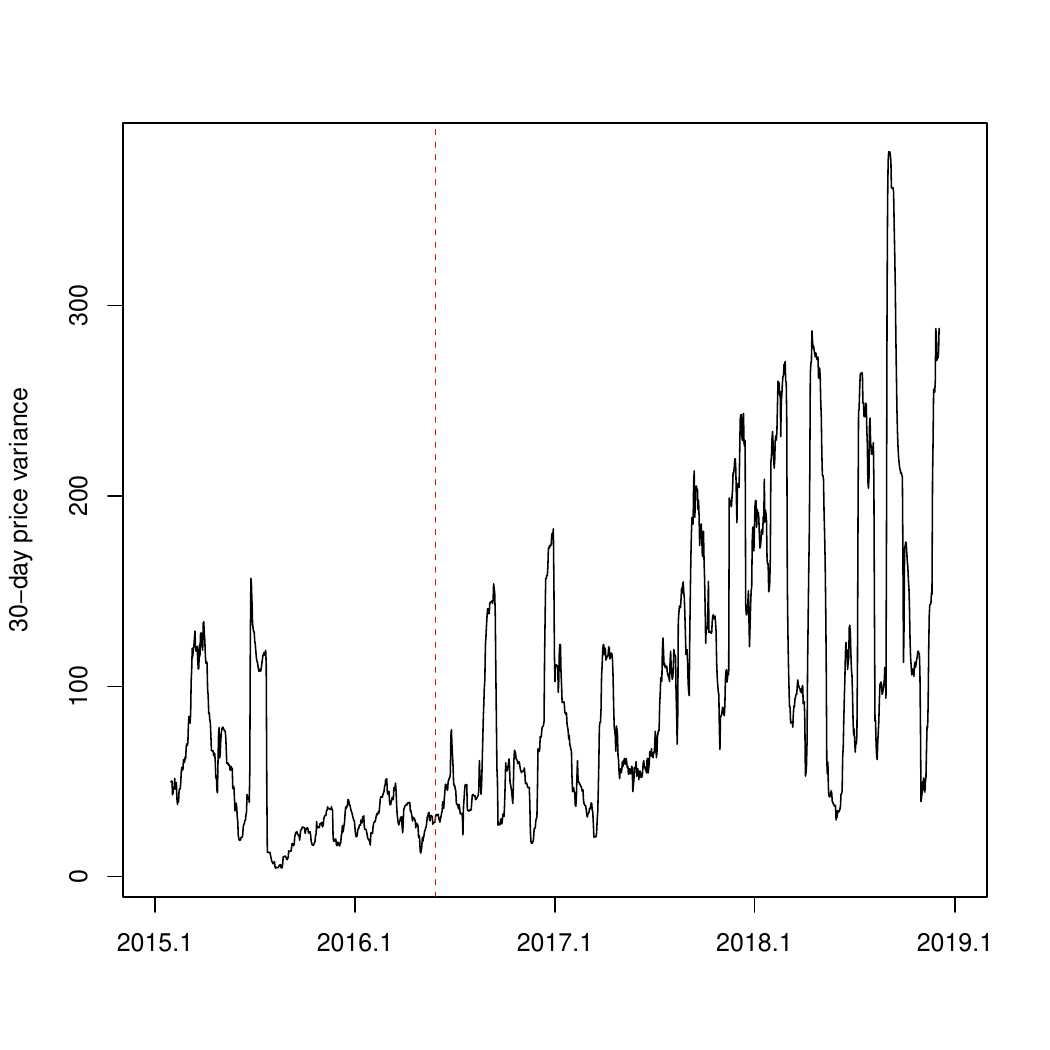}
	\caption{Left: series of the mean daily electricity price in Sicily between 2015 and 2018. Right: 30-day rolling variance of price in the same period. The dashed vertical line is set at May, 28th 2016.}
	\label{prezzi_sicilia}
\end{figure}

This study examines zonal prices in the Italian day-ahead wholesale electricity market from January 1, 2015, to December 31, 2018 \citep{mercatoelettrico2025}. The period covers roughly 18 months before the interconnector was introduced and two and a half years after, allowing analysis of its effects on price dynamics and volatility. The time frame was not extended further because the primary focus is on the impact of the interconnector, and two and a half years is considered sufficient to observe it. Extending the period further could introduce confounding effects from factors unrelated to the new interconnector.\\
Although the original prices had an hourly frequency, here daily volume-weighted spot electricity prices are considered to avoid specific hourly-related effects, which may lead to a less clean ``signal''.\\
The main goal of this study is to identify and quantify the effects of the introduction of the interconnector on price volatility in Sicily. The analysis was also extended to the other five zones to verify that the observed effects are not driven by common factors affecting the entire electricity market.\\
Following empirical evidence, the left panel of Figure \ref{prezzi_sicilia} shows the mean daily price in Sicily for the period 2015-2018, while the right panel depicts the rolling 30-day variance of the price in the same period. The dashed vertical line is placed at the interconnector introduction. It is evident that prices did not decrease after May 2016. In addition, the 30-day rolling variance appears to have increased.\\
These simple graphs suggest that, although prices in Sicily have not decreased, the interconnector may have increased price volatility. 

If price volatility did increase following the introduction of the interconnector, this change is unlikely to have been sudden. Rather, it is conjectured that it began gradually, eventually reaching a new higher level that persists over time. Consequently, a sigmoid-like intervention function is expected, starting at zero around the interconnector's inauguration, with a left asymptote at zero and levelling off at a plateau representing the permanent effect.

\section{The model framework}\label{sec:models}

To describe the dynamics of price and volatility in each market zone, a heteroscedastic additive regression model previously applied in this context is employed \citep{LISI18,BERNARDI20}.\\
Let $p_{t}$ be the daily volume-weighted spot electricity prices for day $t=1,...,T$ and $\mathbf{x}_{t}=(x_{1,t},...,x_{p,t})$ a vector of $p$-covariates referred to time $t$. According to our model, the dynamics of $p_{t}$ is given by:
\begin{equation}\label{eq.model}
	\begin{split}
		p_{t} & = \mu(\mathbf{x}_{t}) + \sigma_{t}  z_{t},\\
		\vspace*{-1em} 
		& = \mu(\mathbf{x}_{t}) + \epsilon_{t},
	\end{split}
\end{equation}
\noindent where $z_{t} \overset{\text{i.i.d.}}{\sim} N(0, 1)$, while $\mu(\mathbf{x}_{t})$ and $\sigma_{t}$ represent the conditional mean and the conditional variance, possibly as a function of the vector of covariates $\mathbf{x}_{t}$, so that:
\begin{equation}
	\text{E}(p_{t}|\mathbf{x}_{t})=\mu(\mathbf{x}_{t}),
\end{equation}
and
\begin{equation}
	\text{Var}(\epsilon_{t}|\mathbf{x}_{t})\equiv\sigma_t^2\equiv\text{E}[(p_{t}-\mu(\mathbf{x}_{t}))^2|\mathbf{x}_{t}]=\text{E}(\epsilon^2_{t}|\mathbf{x}_{t}).
\end{equation}
In this context, also for prices in financial markets, the definition of volatility refers to the conditional variance $\sigma_t^2$ of the residuals of the conditional mean.\\
In particular, the conditional mean is specified with non-parametric additive functions of the covariates $x_{j,t}$:
\begin{equation}\label{eq.condmean}
	\mu_{t} \equiv \mu(\mathbf{x}_{t}) = \beta_0 +\sum_{j=1}^p f_j (x_{j,t}, \lambda_j), 
\end{equation}
where $f_j$ are spline functions (or other smoothers) describing the effect that each variable belonging to $\mathbf{x}_{t}$ has on $\mu_{t}$ conditionally on the values of all other variables and $\lambda_j$ are suitable smoothing parameters.\\ 
For the conditional variance, two different specifications are considered. The first is a parametric GARCH(1,1) with an additive intervention function, $Int_t$, devoted to describe the effect of the new interconnector. In this case, the measure of volatility is given by:
\begin{equation}
	\sigma^2_{t}=\omega+\alpha \ \epsilon^2_{t-1} +
		\beta \sigma^2_{t-1} + Int_t
	\label{garch1_vol}
\end{equation}
where $\omega$, $\alpha$ and $\beta$ are the GARCH parameters.\\ 
The second form of conditional variance is non-parametric, meaning that any specific functional form is not imposed but leave the data to shape the effect. This can be done by modelling $\sigma_t^2$ through spline functions and has also been done for the conditional mean. In this case, the measure of volatility is given by:
\begin{equation}\label{eq.condvariance}
	\sigma^2_{t} \equiv \sigma^2(\mathbf{x}_{t}) = \gamma_0 +\sum_{j=1}^q h_j (x_{j,t}, \lambda_j),
\end{equation}
where, again, $h_j$ are smoothers (spline functions) describing the effect that the variables in $\mathbf{x}_{t}$ have on the conditional variance and $\lambda_j$ are smoothing parameters.

\subsection{The identified models}
\begin{figure}[H]
	\centering
	\includegraphics[width=6.6cm, height=5cm]{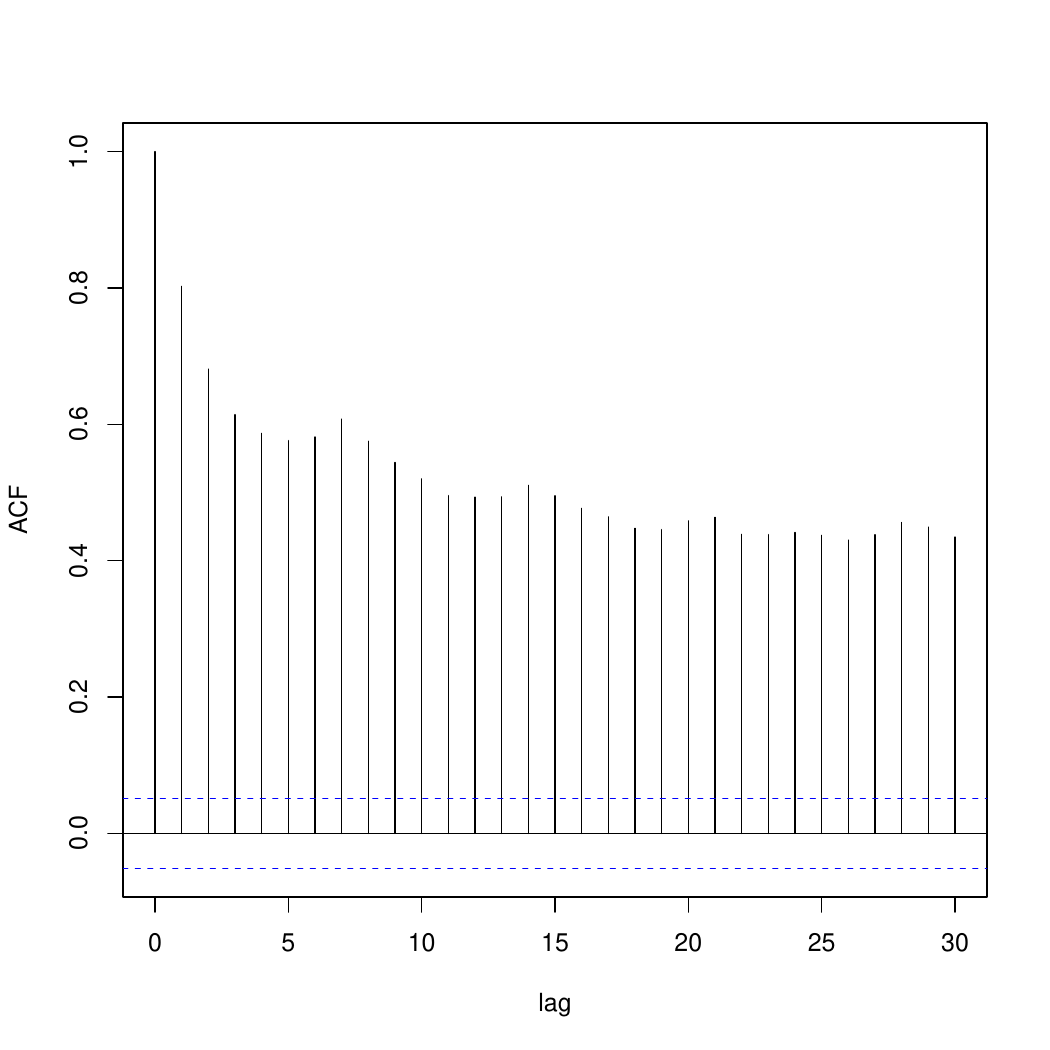}
	\includegraphics[width=6.6
	cm, height=5cm]{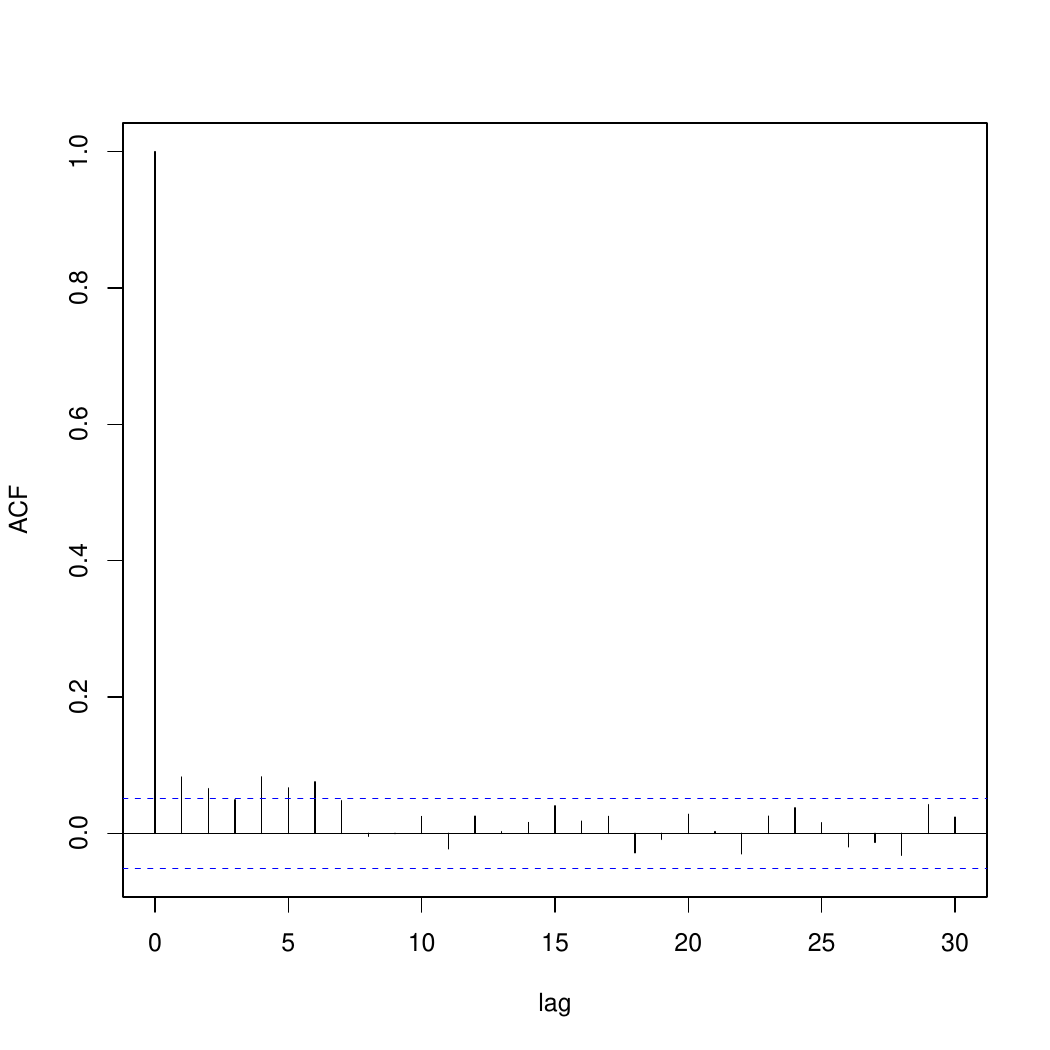}
	\caption{Left: ACF of the original price in Sicily. Right: ACF of the model residuals.}
	\label{res_modello_media}
\end{figure}

\begin{figure}[H]
	\centering
	\includegraphics[width=10cm, height=8cm]{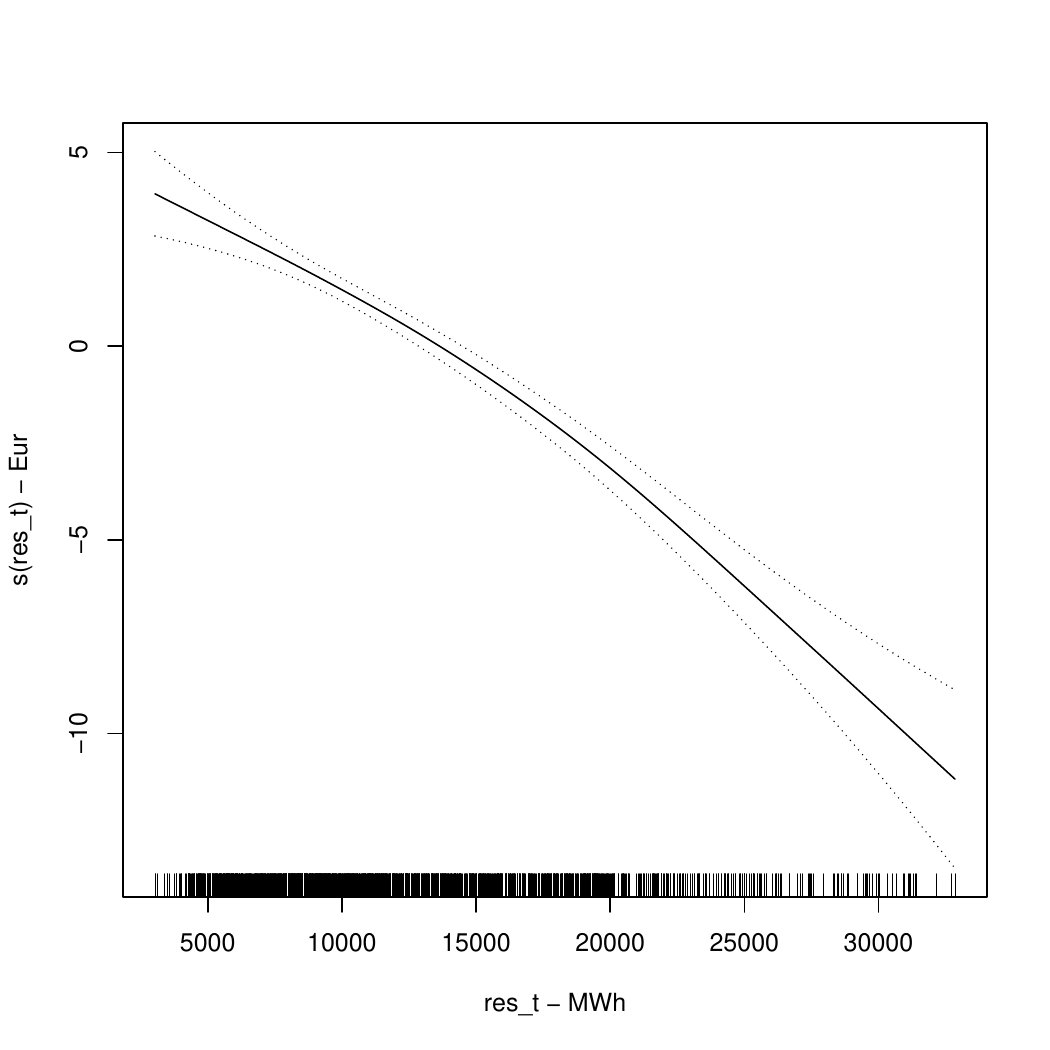}
	\caption{The estimated effect of the RES production (MWh) on the price (Eur) according to model (\ref{mod:mean}).}
	\label{fer_modello}
\end{figure}

The previous section presented the general expressions of the model. In the current application, expression (\ref{eq.condmean}) has been specified as follows:
\begin{equation}
	\label{mod:mean}
	\begin{split}
		\mu_{t}=&\beta_0+f_1(\texttt{trend}_{t}, \lambda_1)+f_2(\texttt{dayyear}_{t},\lambda_2)+f_3(\texttt{dayweek}_{t},\lambda_3)+\delta_1\texttt{bank}_{t}\vspace*{-0.5em}\\
		&+f_4(p_{t-1},\lambda_4)+f_5(p_{t-7} \lambda_5)+f_6(\texttt{res}_{t},\lambda_7),
	\end{split}
	\vspace*{-0.5em}
\end{equation}
where:
\begin{itemize}
	\item $\texttt{trend}_{t}$ is the long-term dynamics on day $t$ and is represented by the sequence $1,2,...,n$;
	\item $\texttt{dayyear}_{t}$ represents the yearly periodicity of the data. It is a vector that repeats the sequence cyclically $1,...,365$;
	\item $\texttt{dayweek}_{t}$ represents the weekly periodicity of the data. It is a vector that cyclically repeats the sequence $1,...,7$;
	\item $\texttt{bank}_{t}$ is a dummy variable for bank holidays;
	\item $p_{t-1}$ and $p_{t-7}$ are the electricity price levels one day and seven days earlier;
	\item $\texttt{res}_{t}$ is the daily generation of renewable energy in MWh.
\end{itemize}
Price levels are modelled as functions of past values, calendar variables that capture long-term dynamics, cyclical components, bank holidays, and the level of renewable energy production. The model is estimated using the backfitting algorithm \citep{Hastie_etal_2001}.
Figure \ref{res_modello_media} shows the ACFs of the original data (left) and of the model's residuals for Sicily (right), which are also representative of the other zones. Although some correlations are still significant, the global correlation structure has been taken into account. Moreover, no autocorrelation exceeds the value of $0.08$ so that the estimated model is satisfactory.\\
Figure \ref{fer_modello} illustrates the estimated relationship between the production of renewable energy (MWh) and the price in Sicily (Eur). The relationship is inverse, indicating that prices decrease as RES production increases. This serves as the initial consideration of renewable energy production in the analysis.\\ 
To model the conditional variance, the residuals of the conditional mean model (\ref{fer_modello}) are used as the starting point.
In both specifications of the conditional variance, an intervention function is introduced to capture the effects of activating the "Sorgente-Rizziconi" interconnector. A \emph{priori}, it is expected that the new interconnection leads to an increase or decrease in price volatility around the inauguration day ($t_0$) and to a plateau within a couple of years. After this period, any further volatility fluctuations are assumed to be not related to the introduction of the interconnector.\\
For these reasons, the intervention function in the parametric form of the conditional variance is defined using a logistic specification, so that expression (\ref{garch1_vol}) becomes:
\begin{equation}
	\begin{split}
		\sigma^2_{t}=&\omega+\alpha \ \epsilon^2_{t-1} +
		\beta \sigma^2_{t-1} + I(t_0) \cdot \frac{a}{1+b \cdot exp(-c \cdot t)}
	\end{split}
	\label{garch2_vol}
\end{equation}
where $I(t_0)$ is an indicator function such that $I(t_0)=1$ if $t \geq t_0$ and $0$ otherwise, $a$, $b$, and $c$ are the parameters of the logistic function. In this context, the important parameters are $a$ and $c$. In fact, $a$ defines the difference between the two asymptotes of the logistic curve and therefore quantifies the final effect on the conditional variance of the introduction of the interconnector. The parameter $c$ controls the speed at which the final level is reached: the higher the value, the faster the final level is reached. This model can be considered as an additive variant of the TV-GARCH model \citep{Amado_Terasvirta_2013, Amado_Terasvirta_2017}. The model (\ref{garch2_vol}) is called the GARCH logistic model (GARCH-L), while the general model (\ref{mod:mean})-(\ref{garch2_vol}) is called the semi-parametric model. All the GARCH parameters are estimated by maximising a conditionally Gaussian likelihood.\\
As robustness check of the results obtained with the GARCH-L specification, the conditional variance is also estimated with a non-parametric additive model \citep{Pinilla2021}. This allows the shape of the intervention function to be determined by the data, without requiring a predefined functional form.
More specifically, equation (\ref{eq.condvariance}) can be formulated as follows: 
\begin{equation}
	\begin{split}
		\sigma^2_{t}=&\gamma_0+h_1(\texttt{Int}_{t})+h_2(\texttt{dayyear}_{t})+h_3(\texttt{dayweek}_{t})\\
		&+\gamma_1\texttt{bank}_{t}+h_4(\epsilon^2_{t-1})+h_5(\texttt{MA}_{t-1})+h_6(\texttt{res}_{t}),
	\end{split}
		\label{nonpar_vol}
\end{equation}
where:
\begin{itemize}
	\item $\texttt{Int}_{t}$ is the intervention variable that takes value 0 before the cables' activation and increases linearly after that day;
	\item $\epsilon^2_{t-1}$ denotes the squared residual with one-day of lag;
	\item $\texttt{MA}_{t}$ is a simple 14-day moving average on $\epsilon^2_{t-j}$ for $j=1,...,14$.
\end{itemize}
The expression (\ref{nonpar_vol}) as a proxy of $\sigma^2_{t-1}$ the GARCH dynamics in a non-parametric way through the terms $\epsilon^2_{t-1}$ and $\texttt{MA}_{t-1}$. Furthermore, it incorporates a non-parametric intervention function and calendar effects via $\texttt{dayyear}_{t}$, $\texttt{dayweek}_{t}$, and $\texttt{bank}_{t}$. Finally, the impact of renewable energy production on the conditional variance is captured by $\texttt{res}_{t}$. In this formulation, the entire model is fully non-parametric.

\section{Results}\label{sec:results}

To analyse the effect of the introduction of the interconnector, the average daily electricity prices are considered for Sicily and for the other zones, from 1 January 2015 to 31 December 2018, that is, accounting for around 17 months before and 29 months after the intervention. 
Although the primary focus is on Sicily (SICI), the analysis is extended to the other zones—namely North (NORD), Center-North (CNOR), Center-South (CSUD), South (SUD), and Sardinia (SARD) to allow suitable comparisons. In fact, the effect of the interconnector is expected to affect only Sicily and possibly the adjacent south zones. Moreover, this impact is expected to follow a pattern consistent with that described previously.\\
The estimated intervention functions are evaluated both graphically and by statistical tests.

First, the results produced by the semi-parametric model are analysed, followed by those obtained from the fully non-parametric model to provide a cross-check of the findings. In both cases, the conditional mean is described by the model (\ref{mod:mean}). From the same residuals, models for the conditional variance are then constructed.\\
Furthermore, to ensure a fair comparison among the different zones, the same regressors are included in the analyses of all zones.

To assess the significance of these effects, a t-test on the logistic coefficients of the GARCH-L model was performed, while an ANOVA test was performed for the non-parametric additive model comparing a "complete" model for the conditional variance, which includes the intervention variable, with a "reduced" model that excludes it.

\subsection{Semi-parametric approach: the GARCH-Logistic model}
\begin{figure}[H]
	\centering
	\includegraphics[width=6.6
	cm, height=5cm]{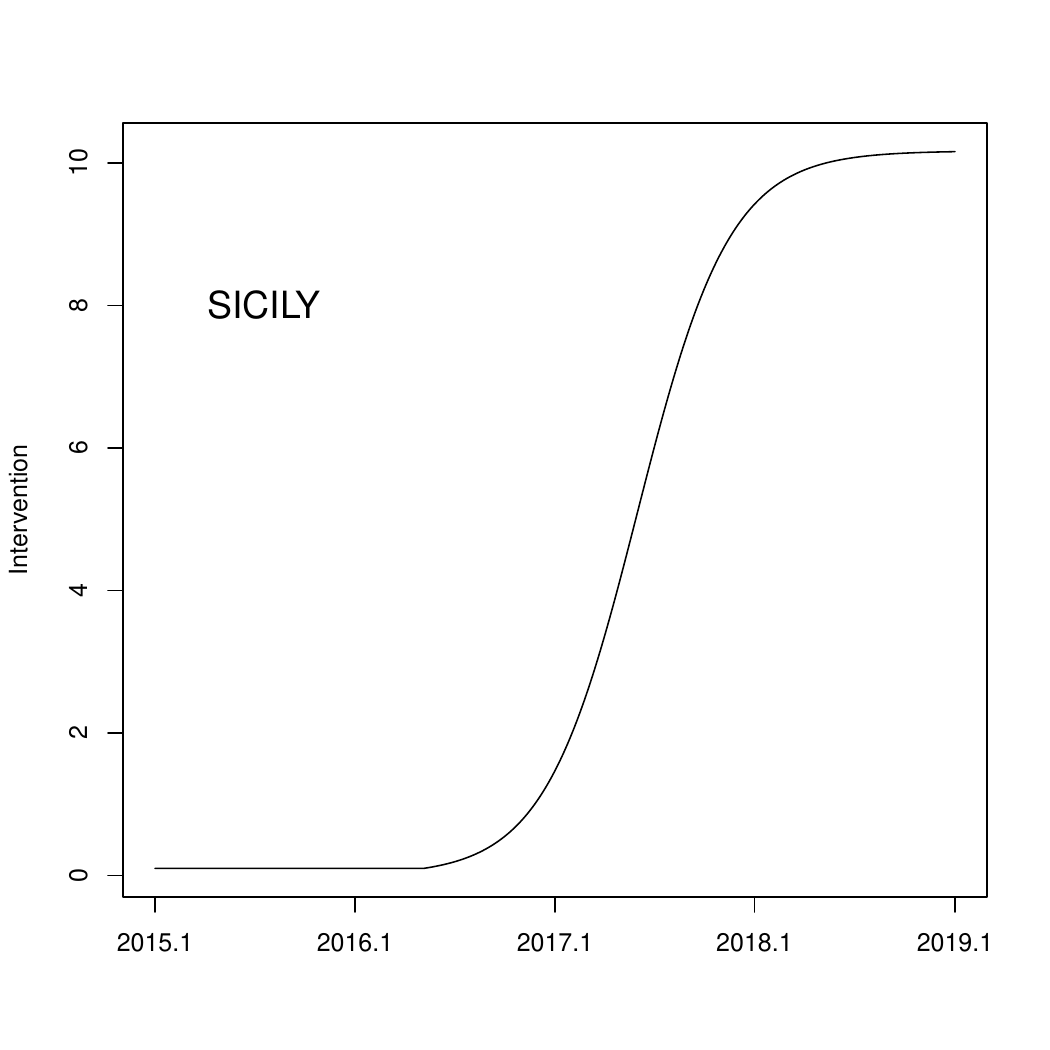}
	\includegraphics[width=6.6
	cm, height=5cm]{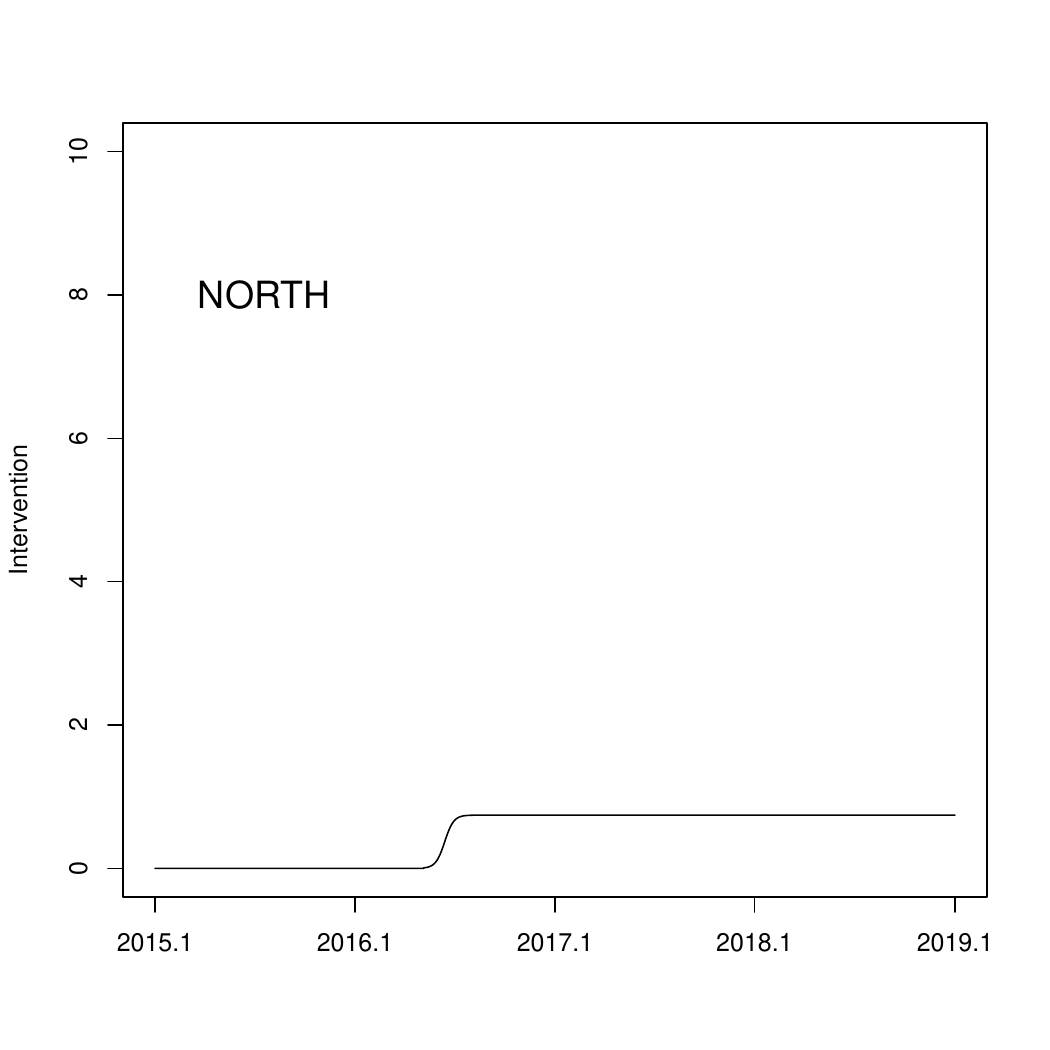}\\
	\includegraphics[width=6.6
	cm, height=5cm]{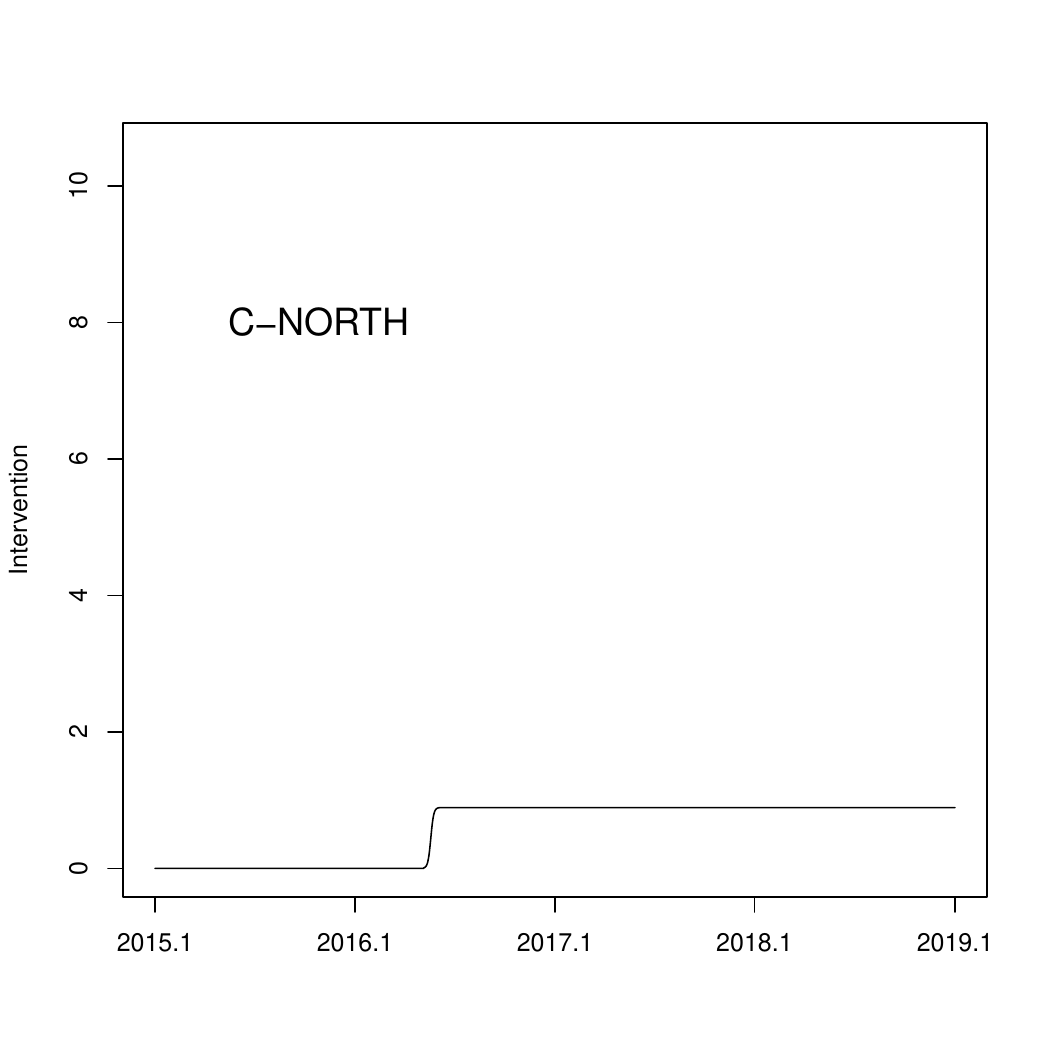}
	\includegraphics[width=6.6
	cm, height=5cm]{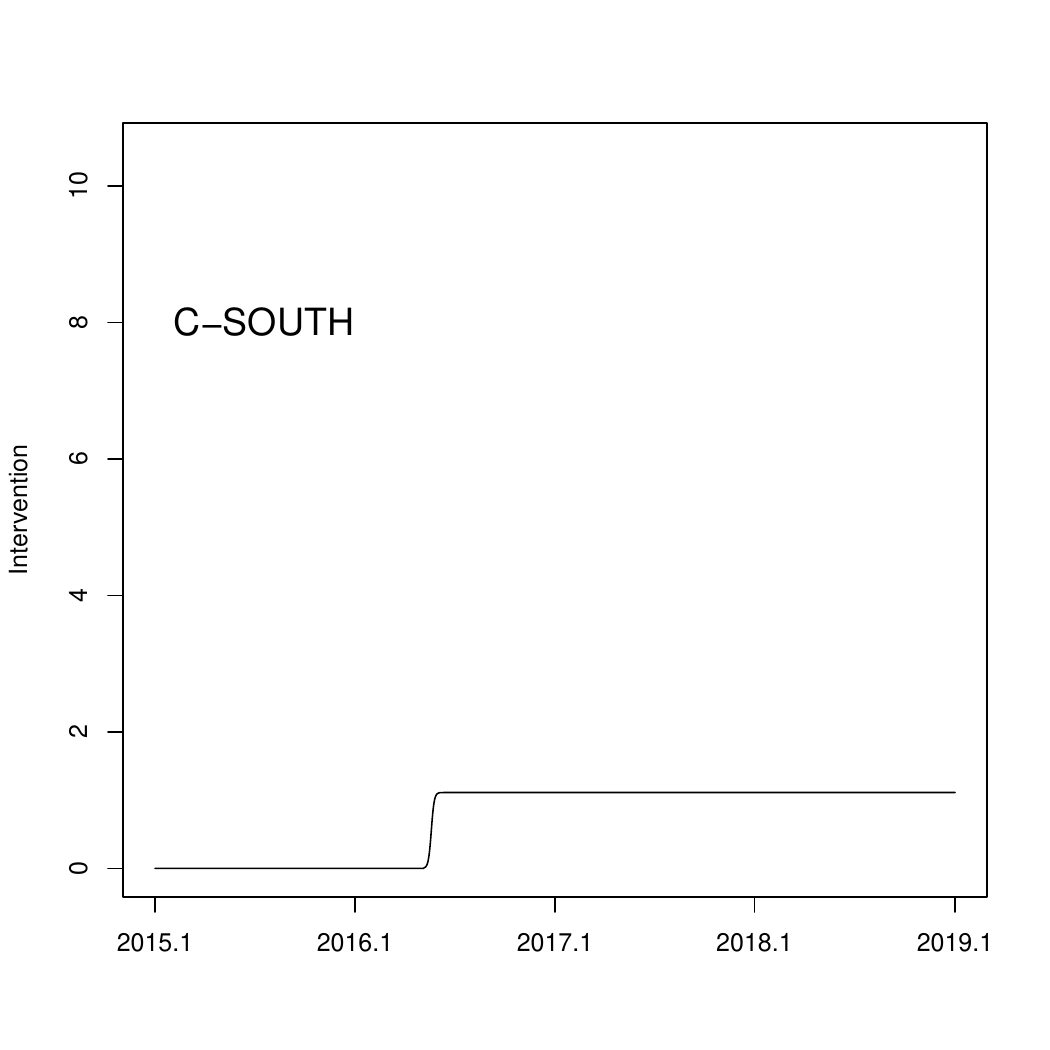}\\
	\includegraphics[width=6.6
	cm, height=5cm]{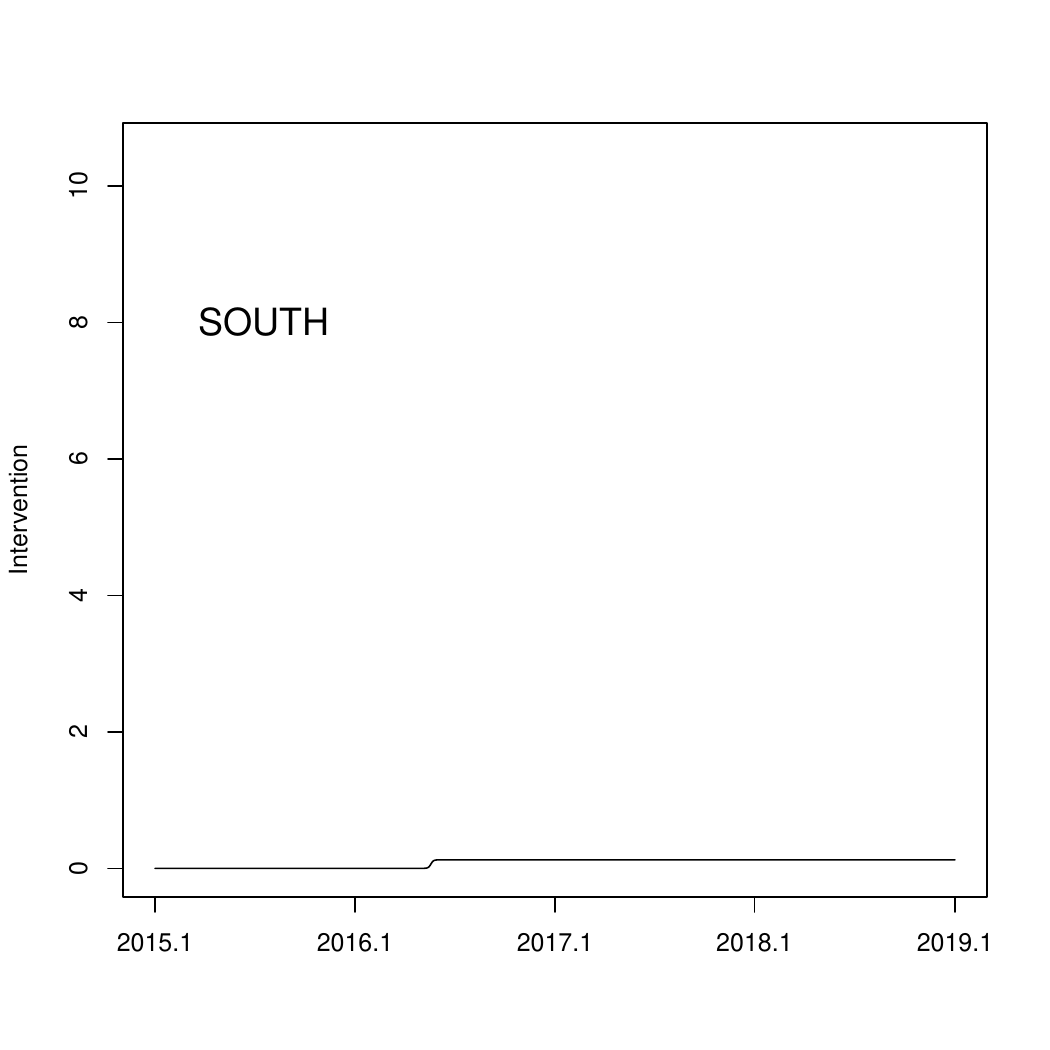}
	\includegraphics[width=6.6
	cm, height=5cm]{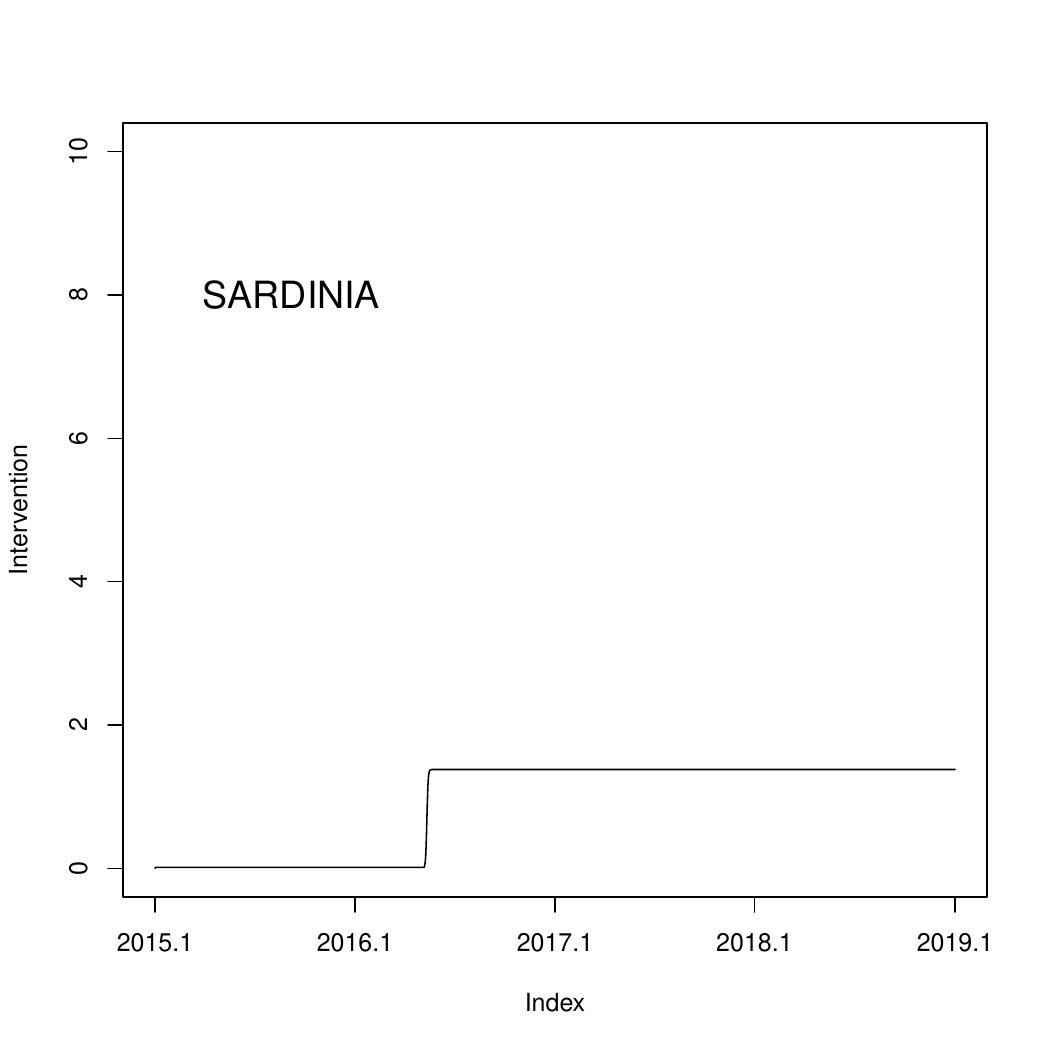}
	\caption{Estimated intervention functions for the six zones on the same scale.}
	\label{interventions}
\end{figure}

\begin{figure}[H]
	\centering
	\includegraphics[width=6.6
	cm, height=5cm]{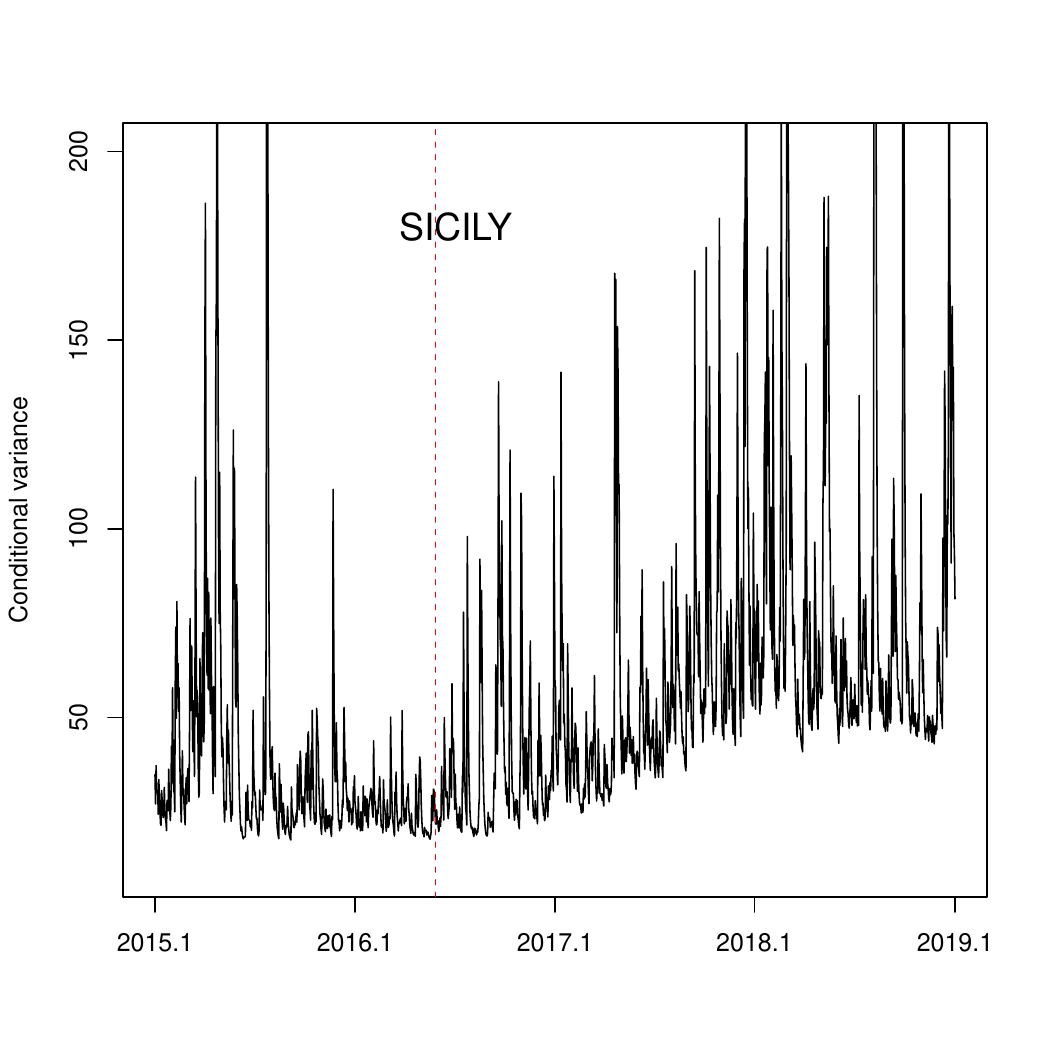}
\includegraphics[width=6.6
cm, height=5cm]{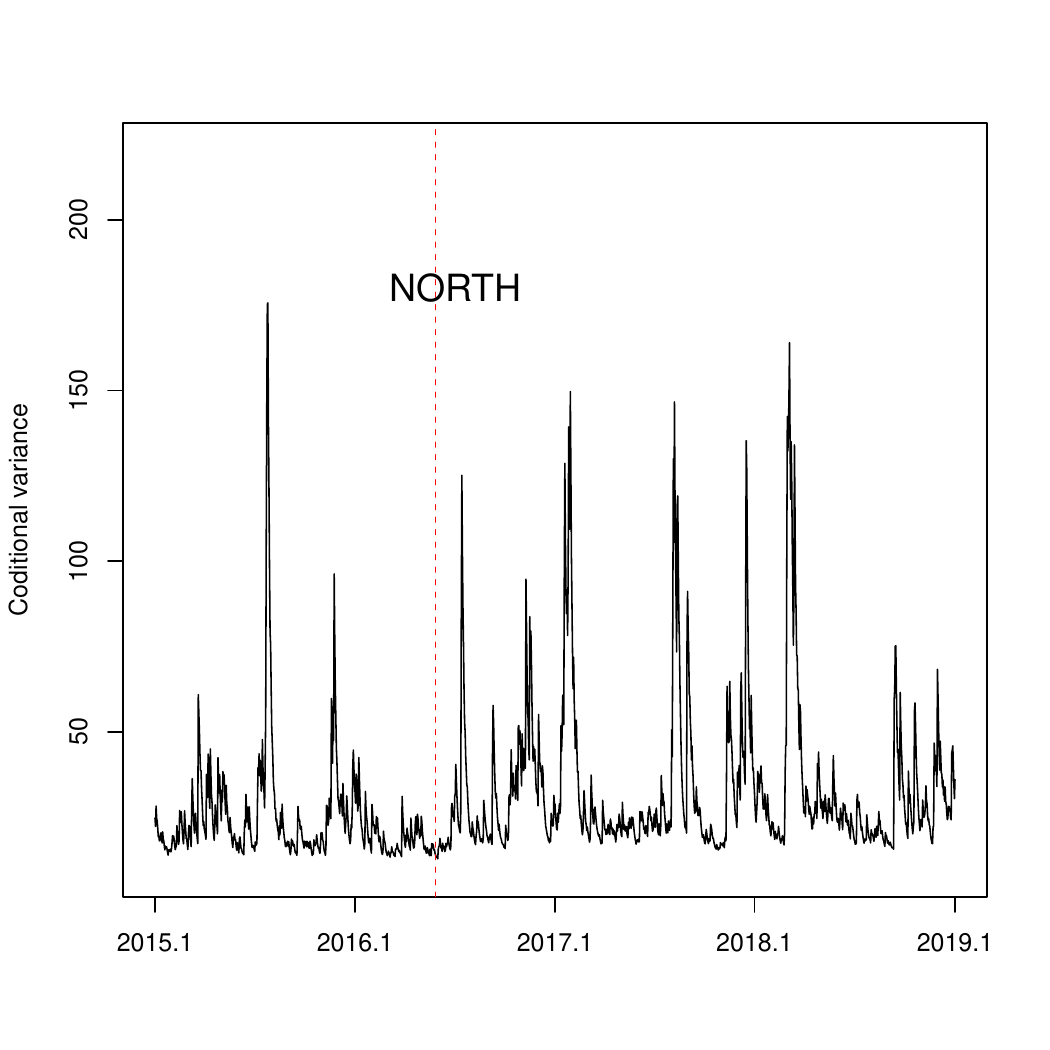}\\
	\includegraphics[width=6.6
cm, height=5cm]{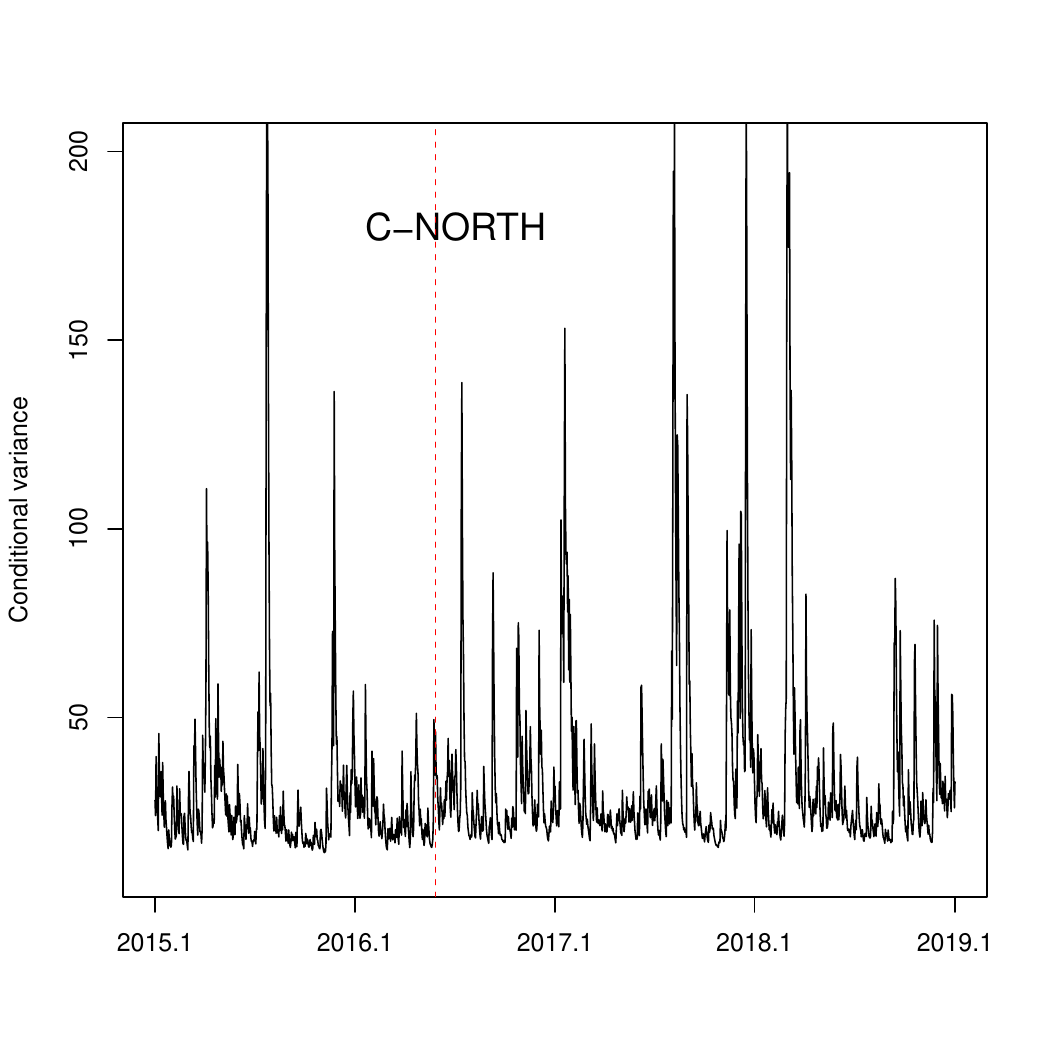}
\includegraphics[width=6.6
cm, height=5cm]{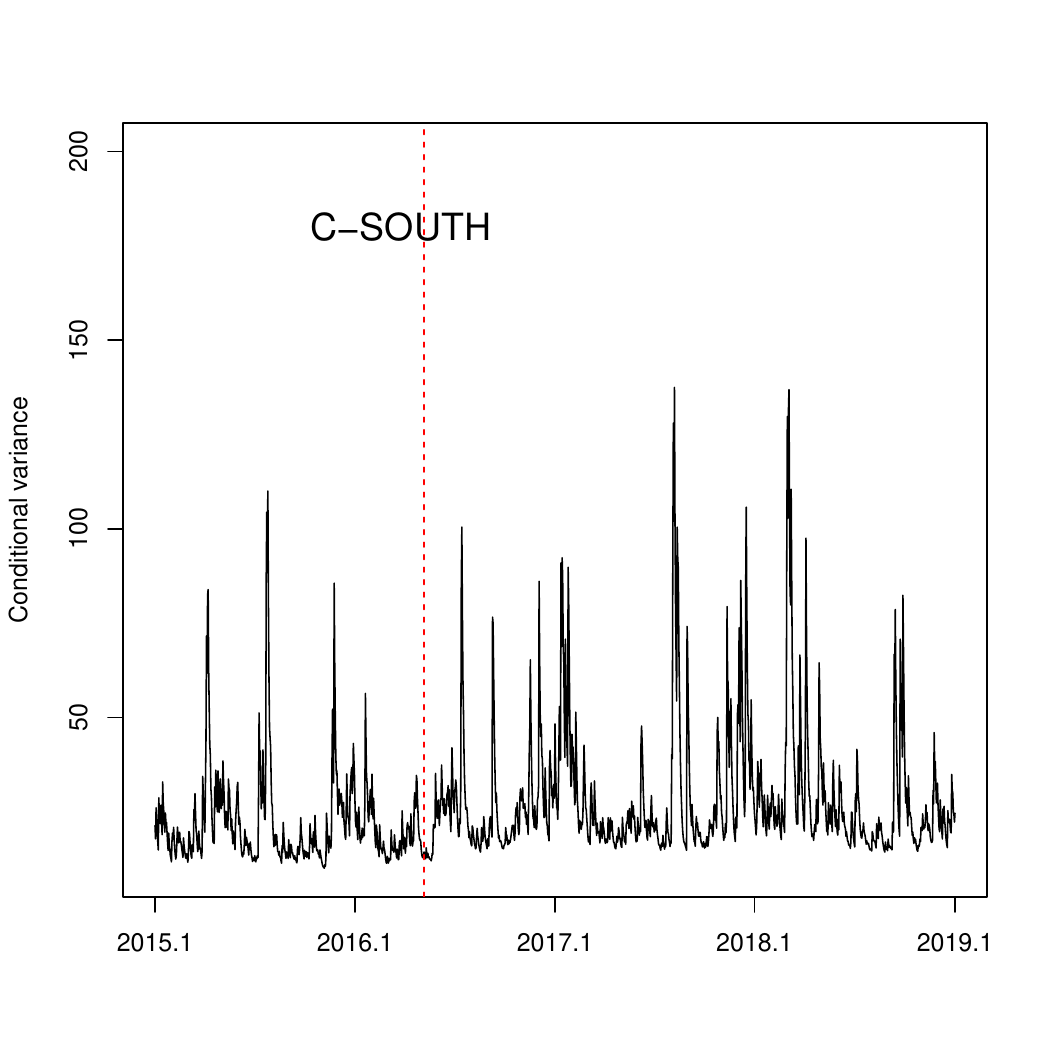}\\
	\includegraphics[width=6.6
cm, height=5cm]{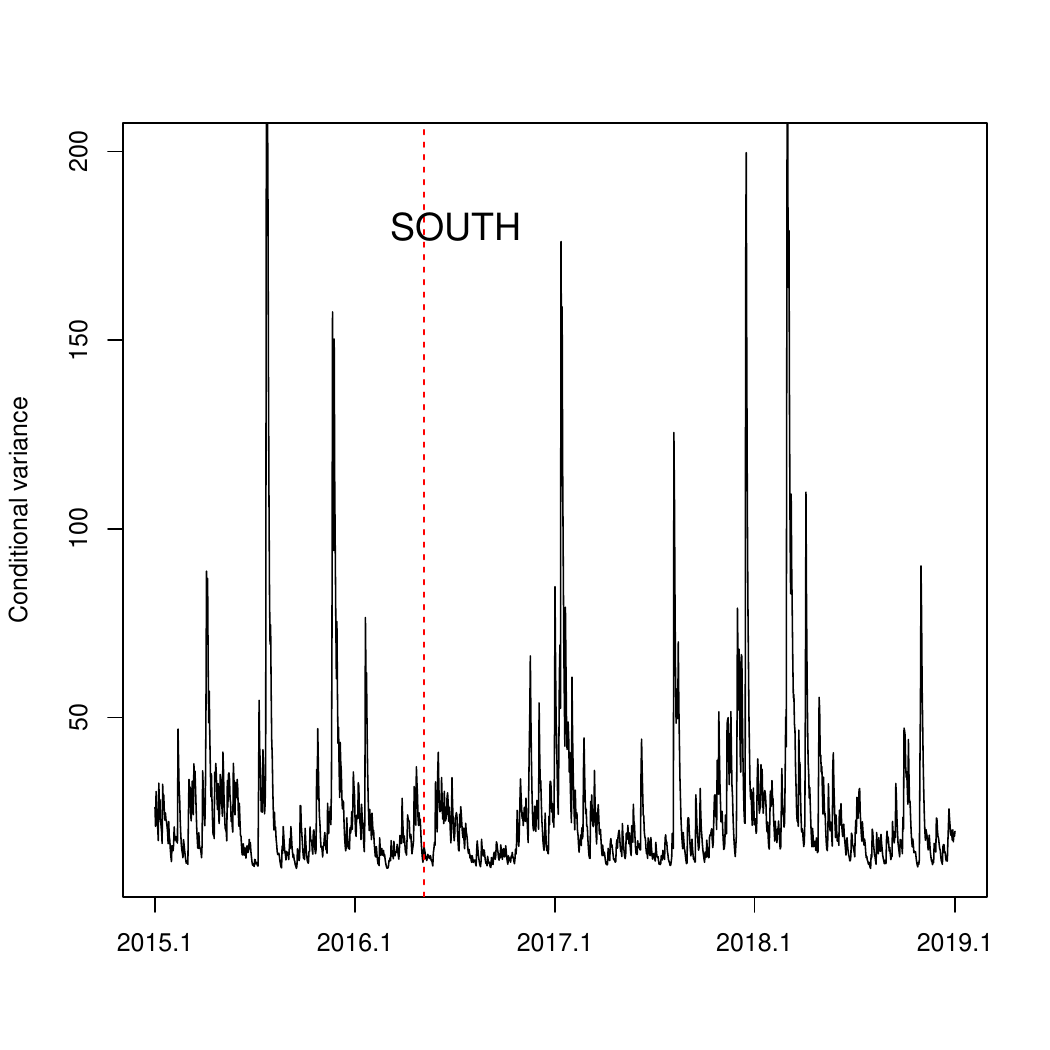}
\includegraphics[width=6.6
cm, height=5cm]{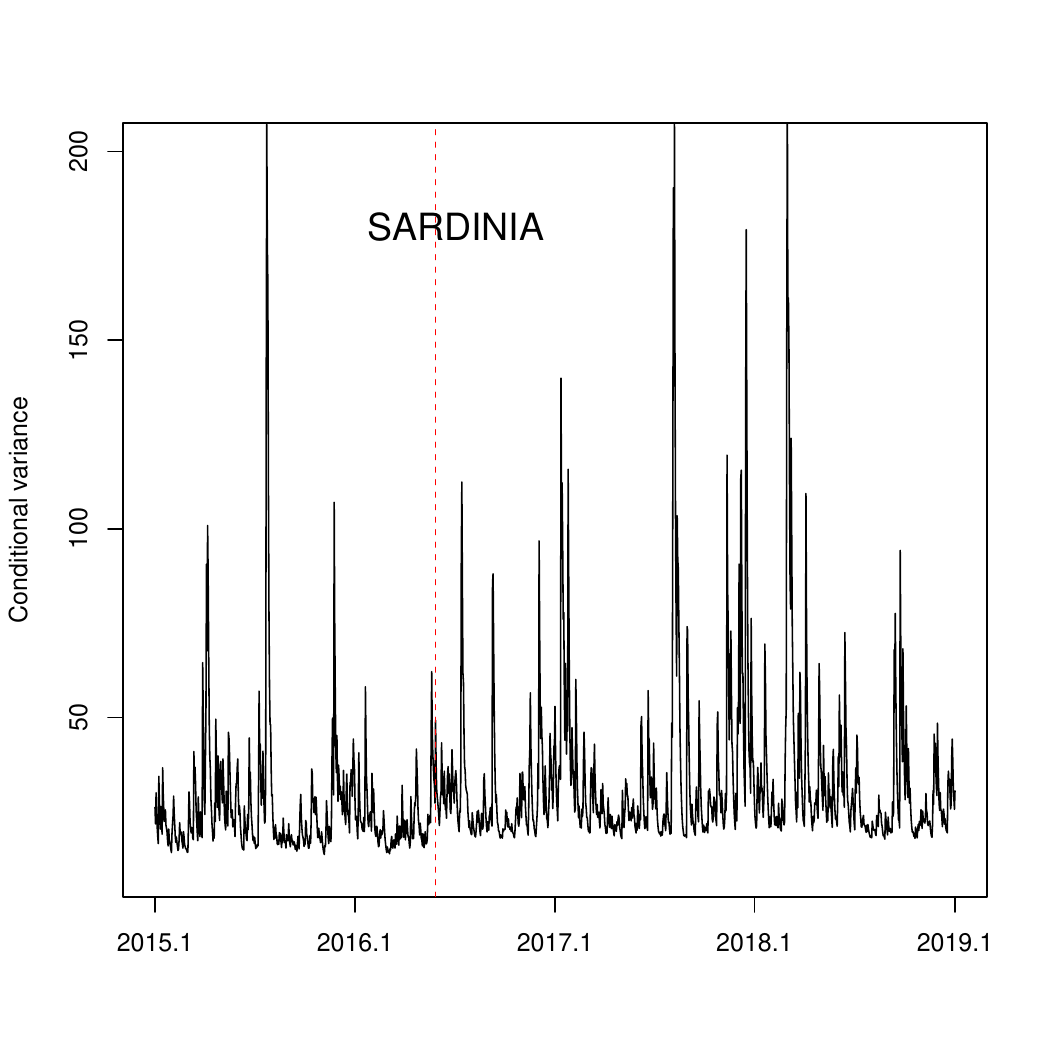}
	\caption{Estimated conditional variance for the six zones on the same scale. The dashed red line is set at 28 May 2016.}
	\label{var-cond}
\end{figure}

First, the semiparametric model (GARCH-L specification) is analysed (\ref{garch2_vol}). \\
Table \ref{garchl_par} lists the estimated parameters of the conditional variance models for all zones. The estimated GARCH parameters, $\hat \omega$, $\hat \alpha$ and $\hat \beta$, are always significant with $\hat \alpha + \hat \beta < 1$, suggesting significant heteroscedasticity within a stationary GARCH(1,1) dynamics. \\
Concerning the parameters of the logistic part, that is, of the intervention function, the key parameters are $a$ and $c$, which quantify the final increase (if any) of the conditional variance implied by the estimated intervention and how quickly the final volatility level is reached. Since the parameter $b$ must be positive, the significance of $\hat b$ is not relevant.\\
Table \ref{garchl_par} shows the estimated parameters and the corresponding p-values for the GARCH-Logistic model and for each zone. It can be observed that for Sicily both $\hat{a}$ and $\hat{c}$ are significant at the $5\%$ level. In particular, the value $\hat{a} = 10.17$ suggests that the introduction of the interconnector increased the base level of price volatility by around 10.2 euros. This represents an increase of almost $29\%$ with respect to the average volatility level before May 2016, which was 35.5 euros.\\
For all other zones, the parameters of the intervention function are not significant, indicating that the new interconnector did not have significant effects on price volatility. This is also confirmed by the values of $\hat{a}$, which are much smaller than in Sicily.\\
Figure \ref{interventions} shows the estimated intervention functions, on the same scale, for all zones. With respect to the model (\ref{garch2_vol}), these can be interpreted as an additive factor that modifies the value of the parameter $\omega$ and thus increases or decreases the base level of volatility. However, it does not affect the peak regime. \\
Only in the case of Sicily is a gradual increase in variability observed, reaching a new, higher level, which corresponds to the conjectured impact. \\
For the other zones, only irrelevant and abrupt changes are observed, attributable to small and not significant variations in the mean of the volatility time series. \\
Figure \ref{var-cond} shows the estimated conditional variance for the different zones with the vertical dashed line set on the day of inauguration. The increase in the base level of the conditional variance is quite apparent for Sicily. In contrast, for all other zones there are no evident changes in the variance apart from, maybe, some increase in the frequency of the peaks.

\begin{table}[H]
	\centering	
	\begin{tabular}{lcccccc}\hline
		Parameters  & $\hat \omega$   & $\hat \alpha$ & $\hat \beta$ & $\hat a$ & $\hat b$ & $\hat c$\\ \hline	
		SICILY  & {\bf 7.391} &  {\bf 0.222} & {\bf 0.565} &  {\bf 10.171} & 101.2 & {\bf 0.012} \\ 
		  & (<0.001) &  (<0.001) & (<0.001) &  (<0.001) & (0.671) & (<0.028) \\
		NORTH  & {\bf 2.224} &  {\bf 0.126} & {\bf 0.784} &  0.742 & 101.0 & 0.112 \\ 
		& (0.001) &  (<0.001) & (<0.001) &  (0.058) & (0.724) & (0.178) \\
		CENTER-NORTH  & {\bf 4.417} &  {\bf 0.189} & {\bf 0.653} &  0.890 & 110.0 & 0.389 \\ 
		& (<0.001) &  (<0.001) & (<0.001) &  (0.163) & (0.870) & (0.504) \\
		CENTER-SOUTH  & {\bf 2.752} &  {\bf 0.165} & {\bf 0.705} &  1.110 & 150.0 & 0.377 \\ 
		& (0.014) &  (<0.001) & (<0.001) &  (0.054) & (0.820) & (0.384) \\
		SOUTH  & {\bf 2.410} &  {\bf 0.193} & {\bf 0.715} &  0.124 & 206.9 & 0.427 \\ 
		& (<0.001) &  (<0.001) & (<0.001) &  (0.773) & (0.970) & (0.821) \\
		SARDINIA  & {\bf 4.420} &  {\bf 0.179} & {\bf 0.651} &  1.378 & 99.1 & 0.881 \\ 
		& (0.002) &  (<0.001) & (<0.001) &  (0.055) & (0.906) & (0.670) \\ \hline
	\end{tabular}
	\caption{Estimated parameters and, in brackets, corresponding p-values for the GARCH-Logistic model and for each zone. Significant estimates are emphasized in bold.}
	\label{garchl_par}
\end{table}

\subsection{Robustness checks}
\begin{figure}[H]
	\centering
	\includegraphics[width=10cm, height=8cm]{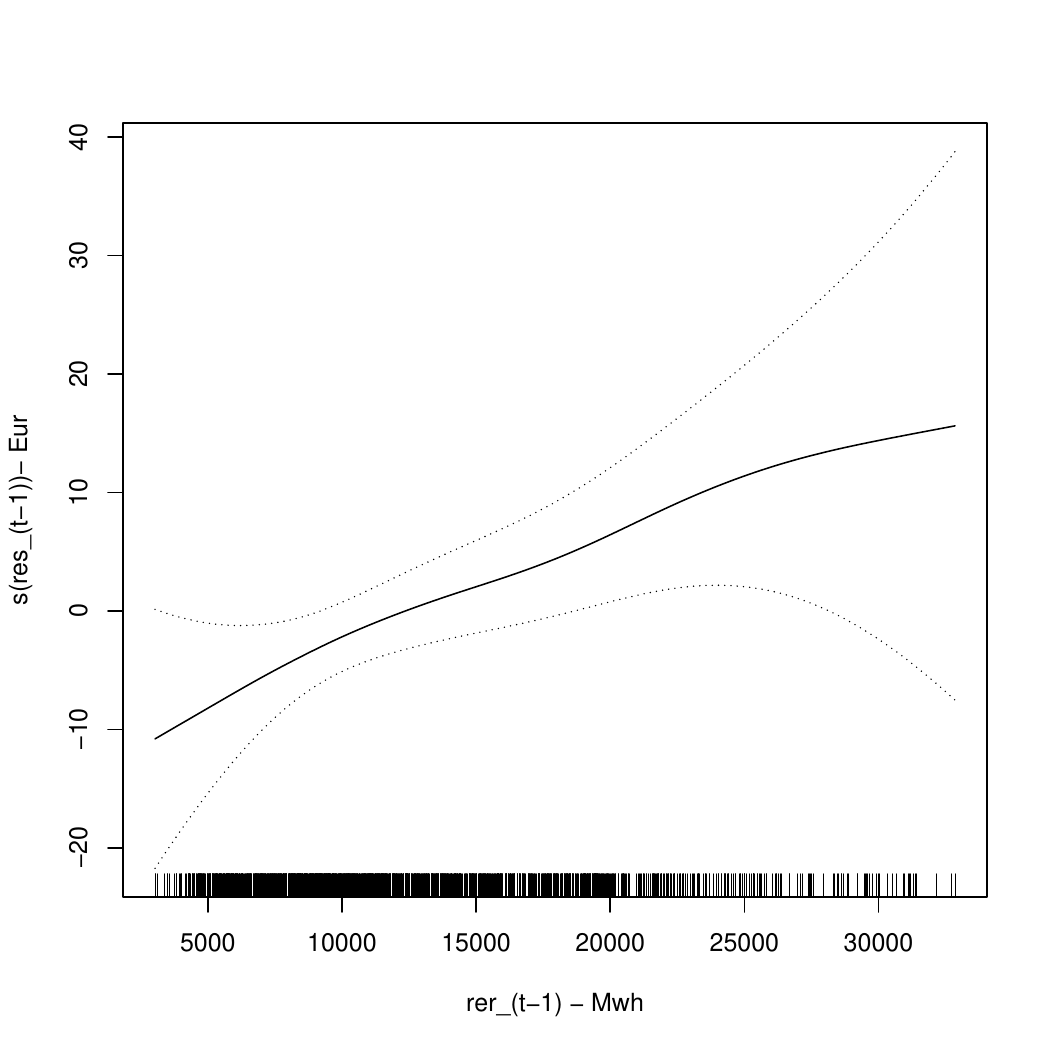}
	\caption{Sicily. Effects of RES (MWh) on the conditional variance (Eur).}
	\label{fer_sicilia}
\end{figure}

\begin{figure}[H]
	\centering
	\includegraphics[width=6.6
	cm, height=5cm]{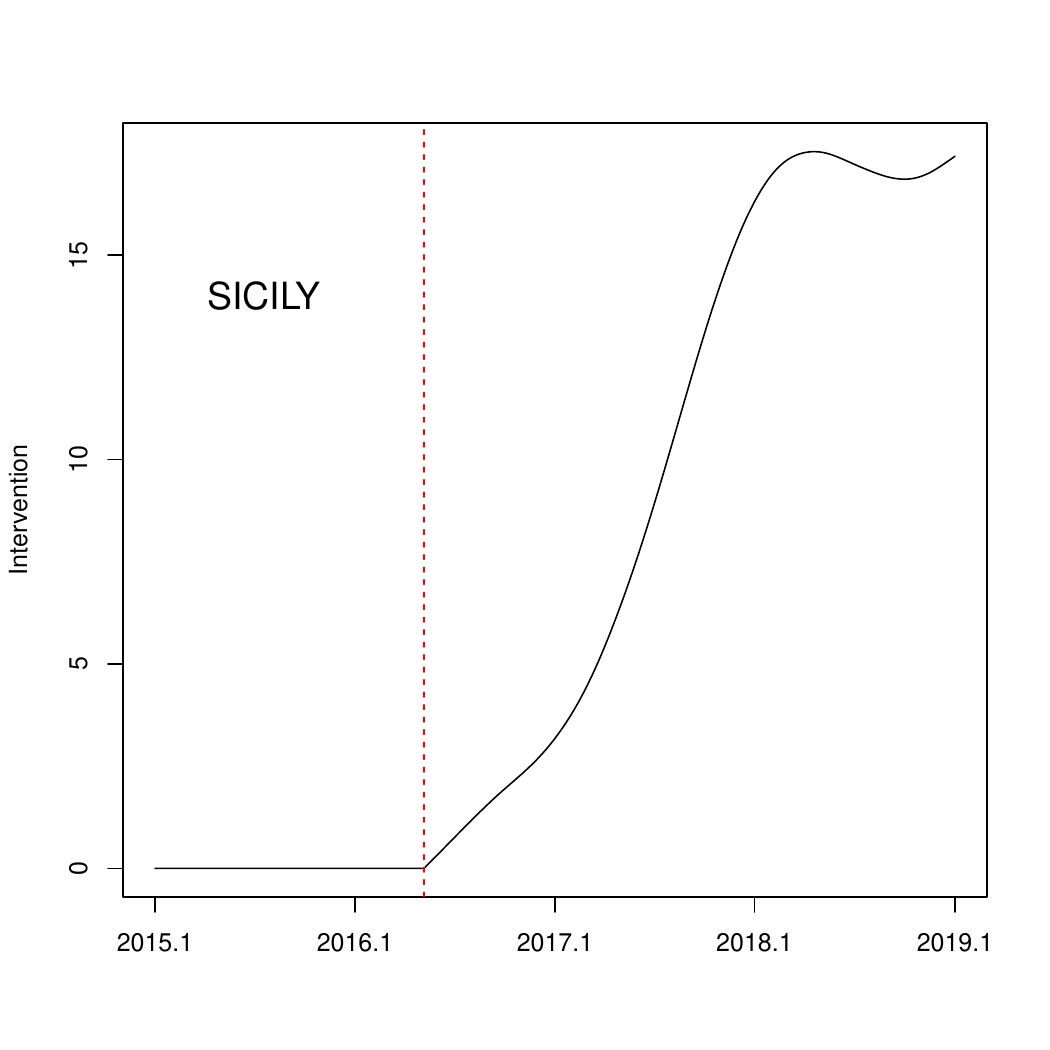}
	\includegraphics[width=6.6
	cm, height=5cm]{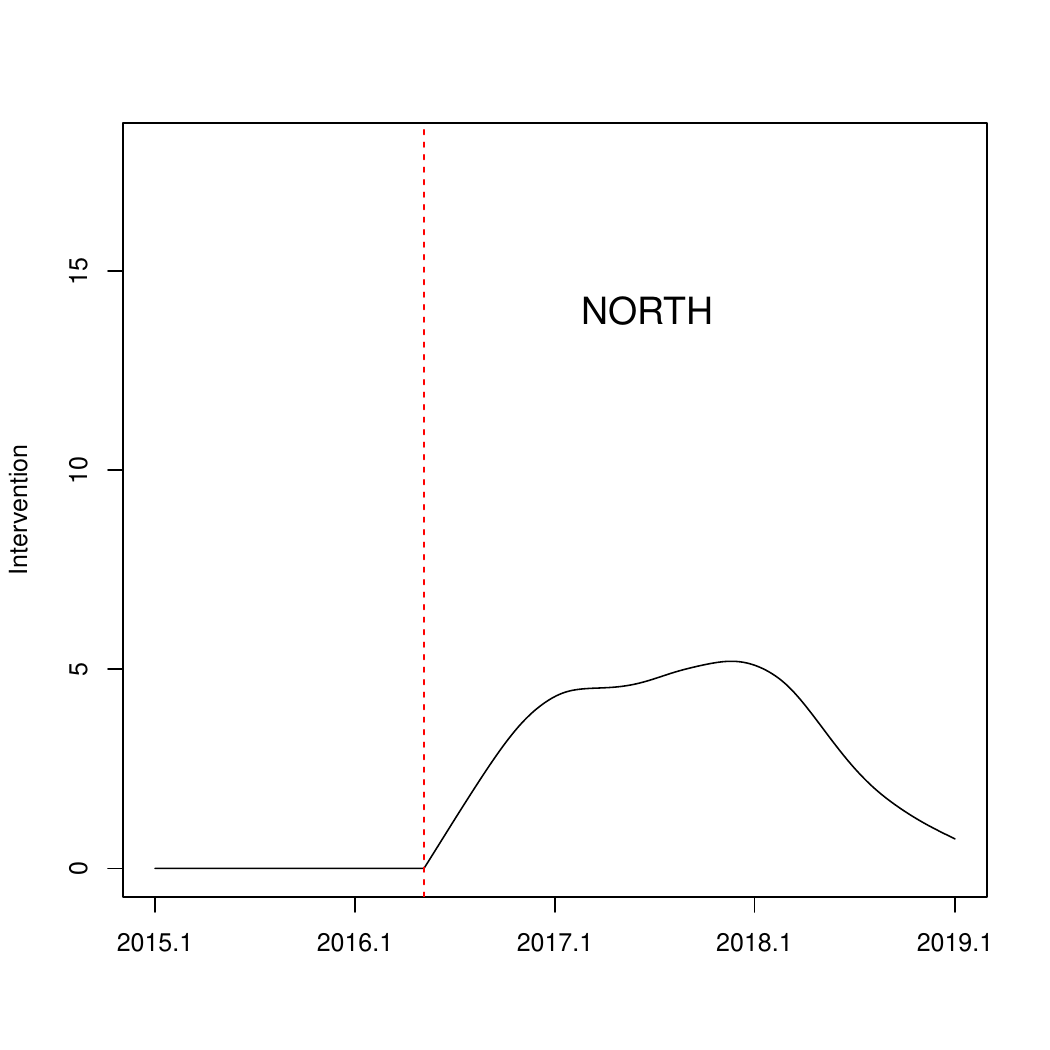}\\
	\includegraphics[width=6.6
	cm, height=5cm]{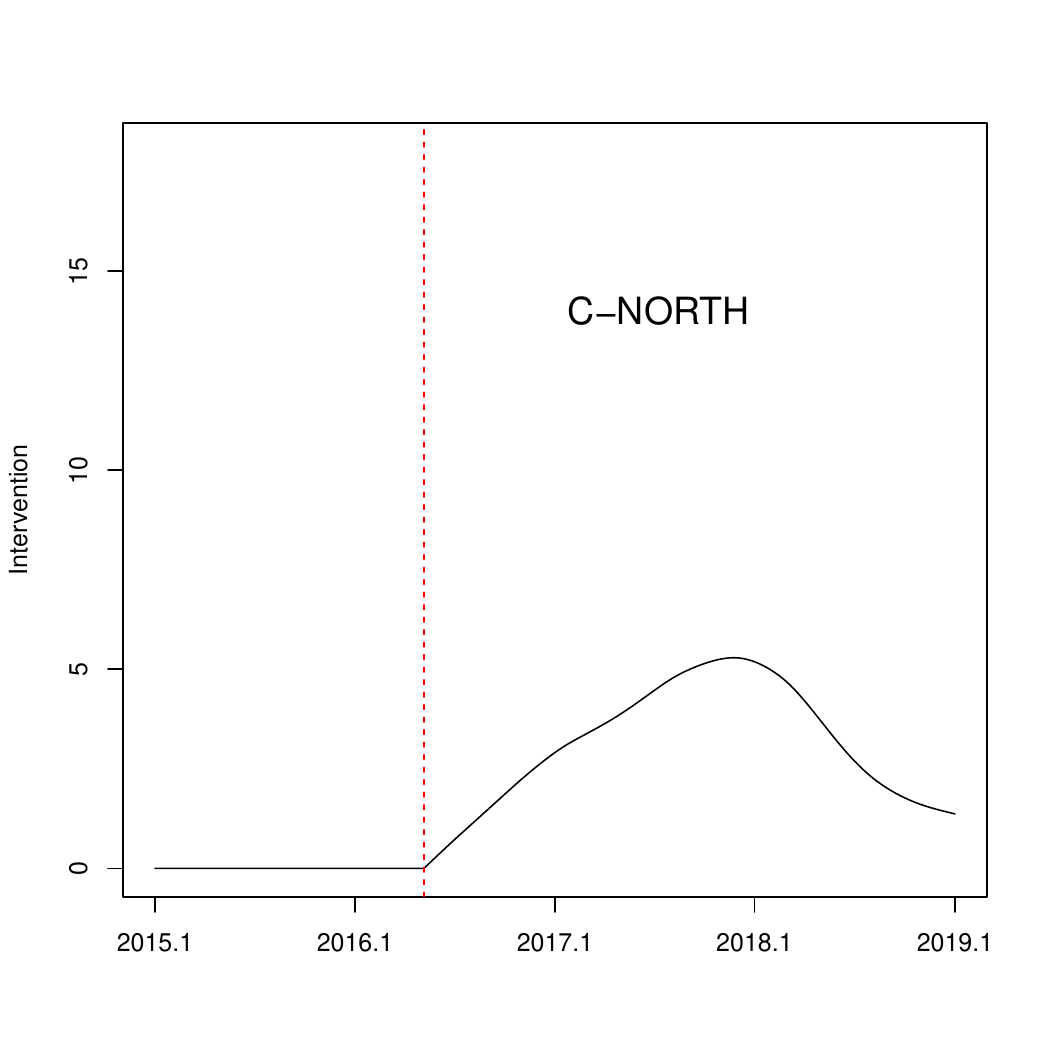}
	\includegraphics[width=6.6
	cm, height=5cm]{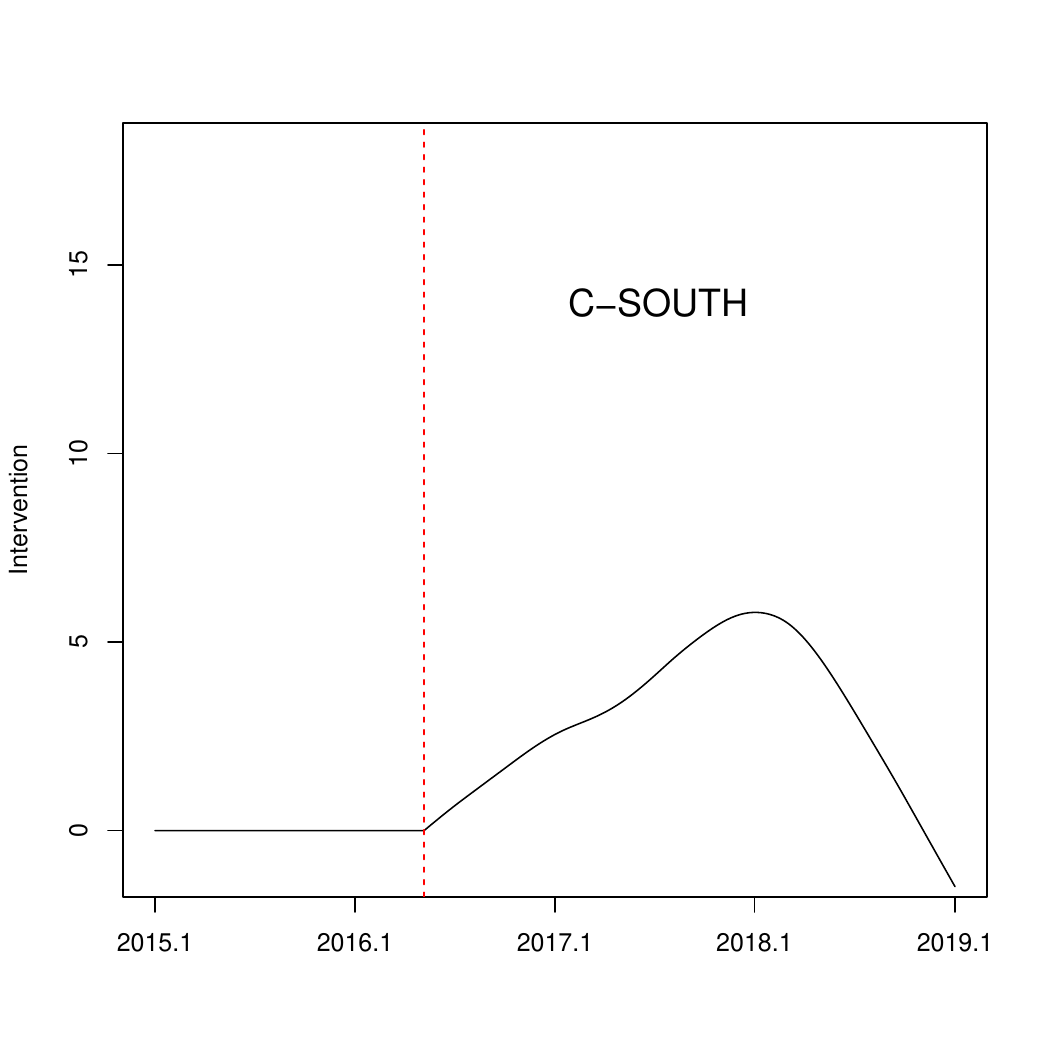}\\
	\includegraphics[width=6.6
	cm, height=5cm]{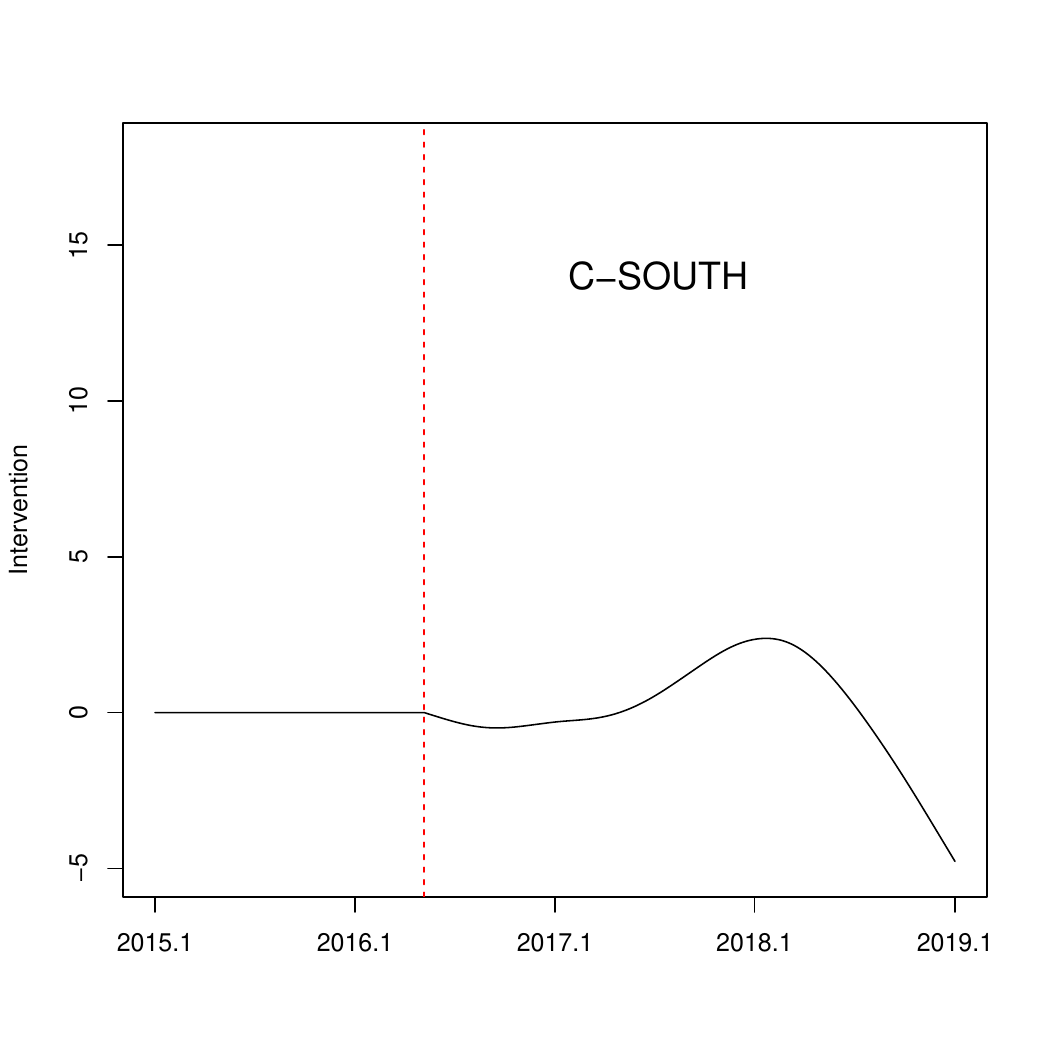}
	\includegraphics[width=6.6
	cm, height=5cm]{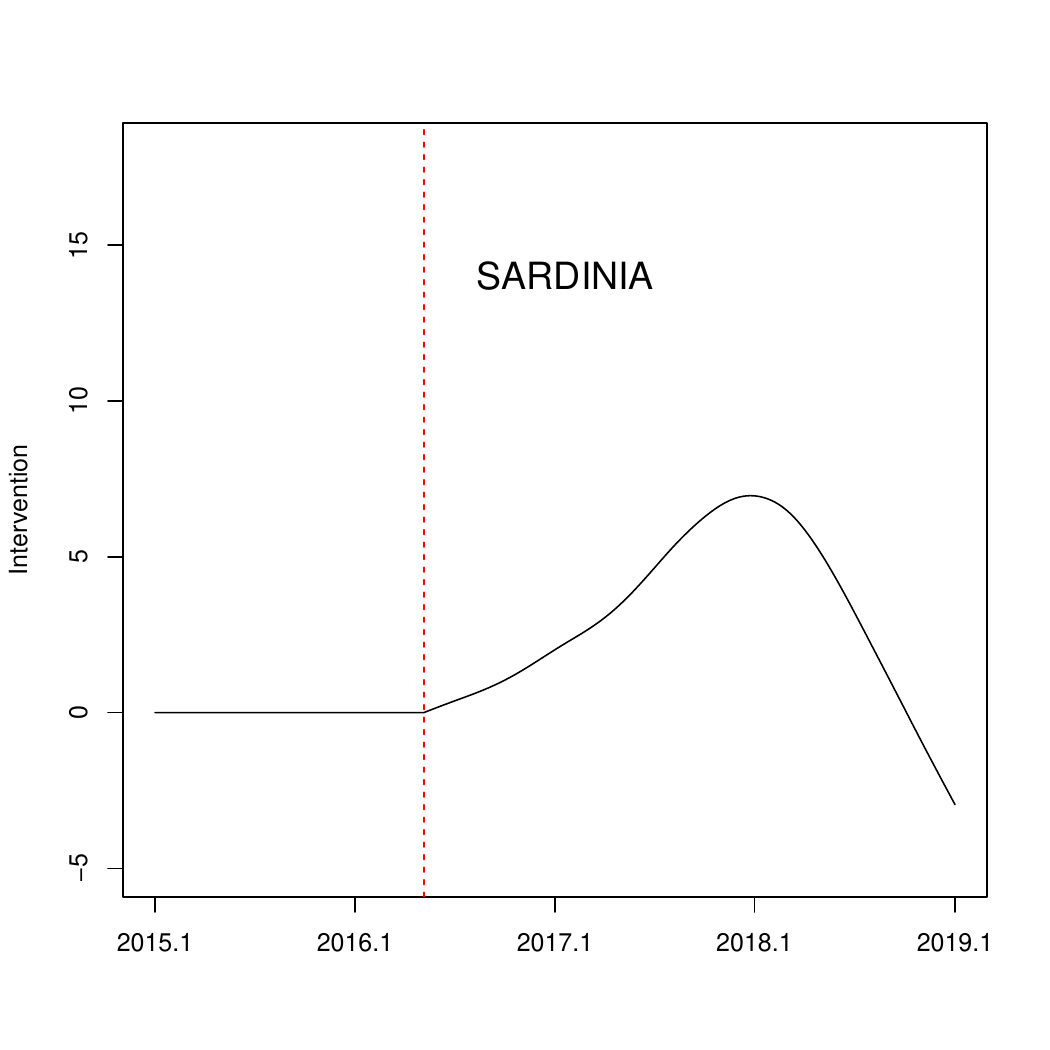}
	\caption{Estimated intervention functions for the six zones on the same scale.}
	\label{nonpar_interv}
\end{figure}

Consider now the fully non-parametric approach for robustness check of the results obtained with GARCH-L specification.\\
An attractive feature of this model is that no functional form is imposed on the intervention function; instead, the data shape the interconnector effect, which may assume any form.\\
Although the results from this model are expected to be less tidy compared to those obtained with the semi-parametric approach, this method is employed to provide additional support to previous findings. \\
Moreover, model (\ref{nonpar_vol}) accounts for variables not present in model (\ref{garch2_vol}), particularly RES production. This is important because an observed increase in volatility might be a consequence of the increase in renewable energy generation during the considered period. In this regard, Figure~\ref{fer_sicilia} illustrates, for Sicily, the impact of RES production (MWh) on the conditional variance (Eur). While in the model for price level (conditional mean), an increase in RES leads to a decrease in price, the volatility shows the opposite effect: consistent with expectations, higher renewable production implies greater uncertainty around prices. \\
Regarding the objective of this work, the relevant issue is that the inclusion of this factor enables the estimation of the intervention function net of the RES production effect, as this influence has already been accounted for.
Furthermore, for the fully nonparametric model (\ref{nonpar_vol}), the estimated intervention functions are considered for the six zones (Figure \ref{nonpar_interv}).  
Unlike in the parametric case, the estimated functions are less well defined. This is not strange because, due to their non-parametric nature, they are much more sensitive to peaks and local variations of the underlying time series.  
However, again the results for Sicily are completely different from those of the other zones. Moreover, despite the fact that no specific functional form is imposed on the intervention, this model uniquely exhibits a sigmoid-like pattern, as expected. In this case, the estimated final volatility increases approximately 17 euros, which is greater than that implied by the parametric model. For the other zones, the interventions are considerably smaller, with patterns that indicate both increasing and decreasing volatility.\\
To assess the significance of the estimated interventions, we apply an ANOVA test considering the ``full'' model (\ref{nonpar_vol}) and a ``reduced'' specification that does not include the intervention component. The underlying hypothesis system is the following
\begin{equation}
	\begin{cases}
		H_0: \text{no significant effect of the intervention variable;} \\
		H_1: \text{the intervention variable has a significant effect.} 
	\end{cases}
\end{equation}

\begin{figure}[H]
	\centering
	\includegraphics[width=6.6
	cm, height=5cm]{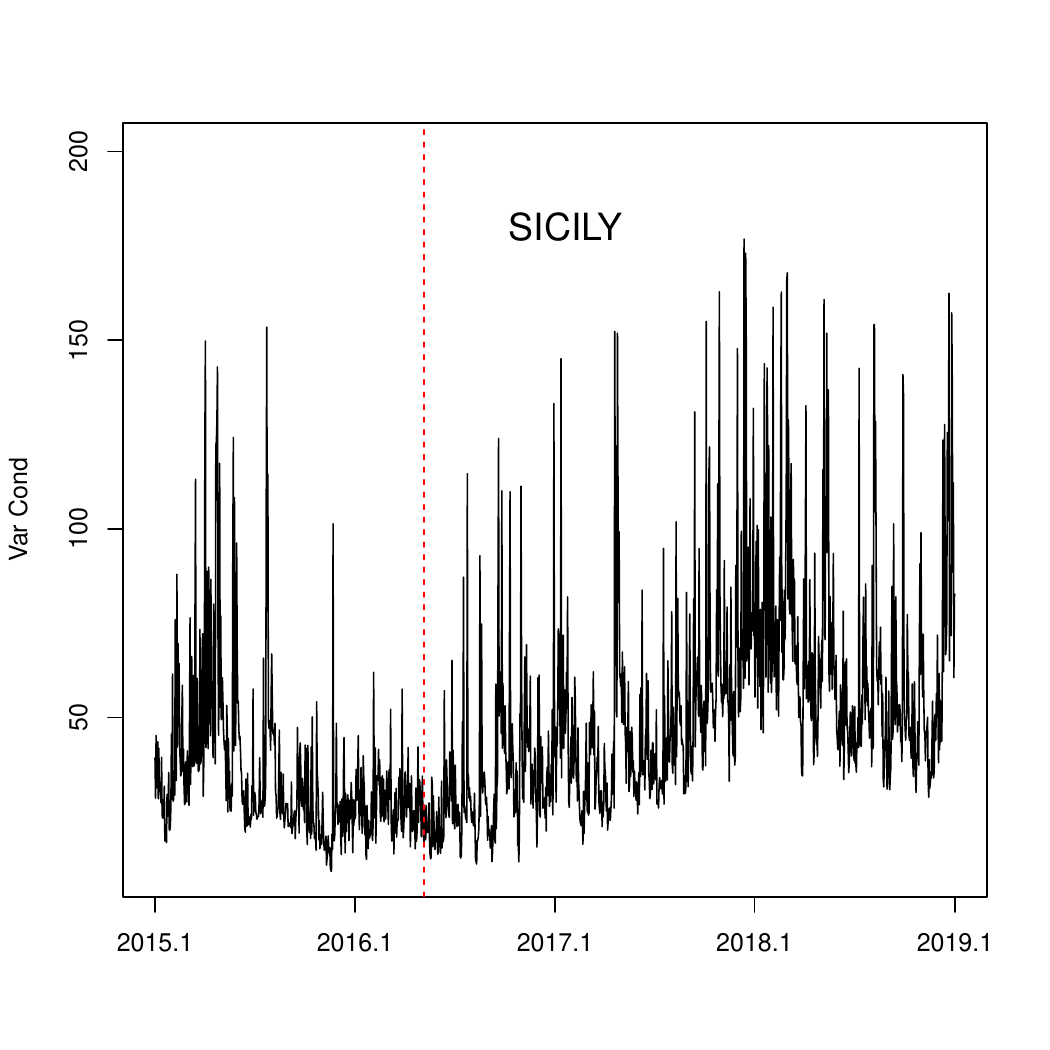}
	\includegraphics[width=6.6
	cm, height=5cm]{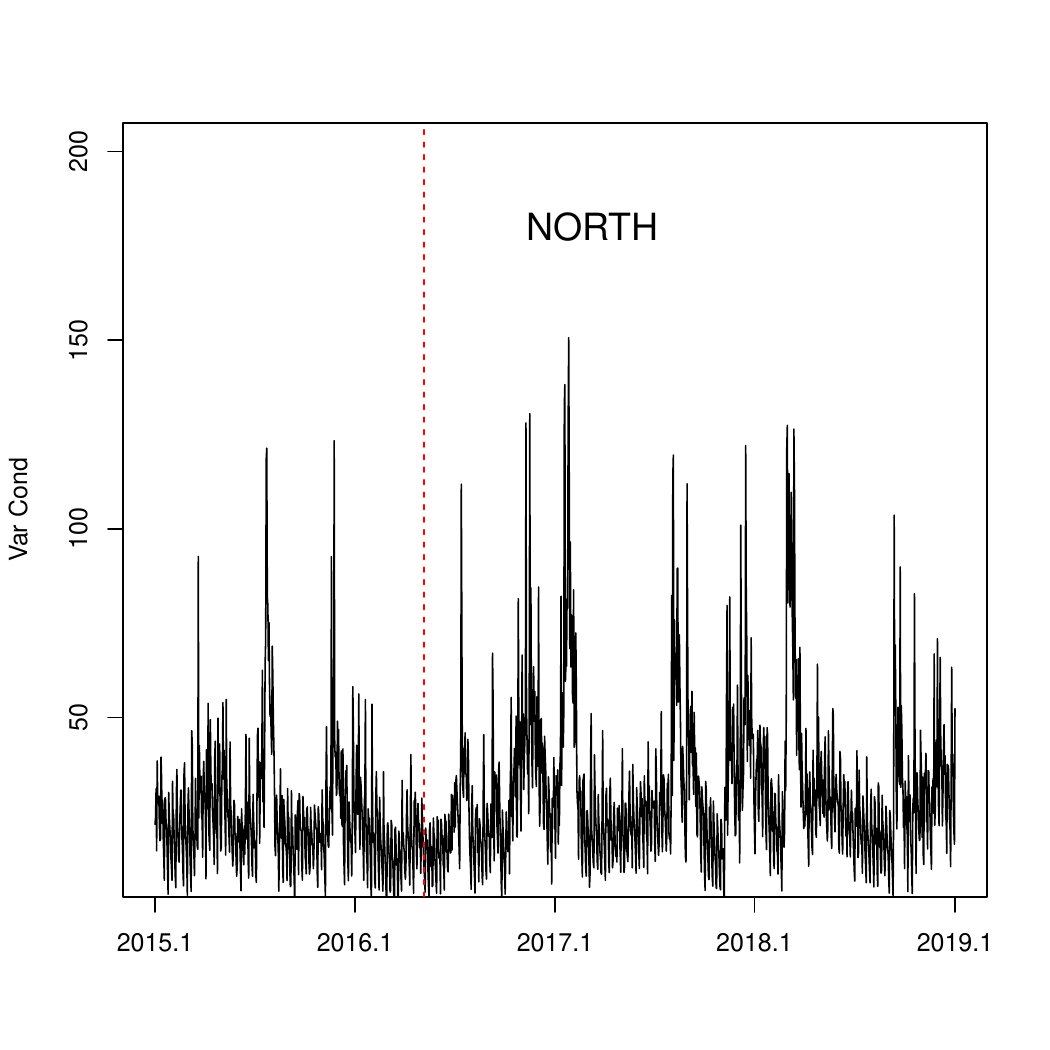}\\
	\includegraphics[width=6.6
	cm, height=5cm]{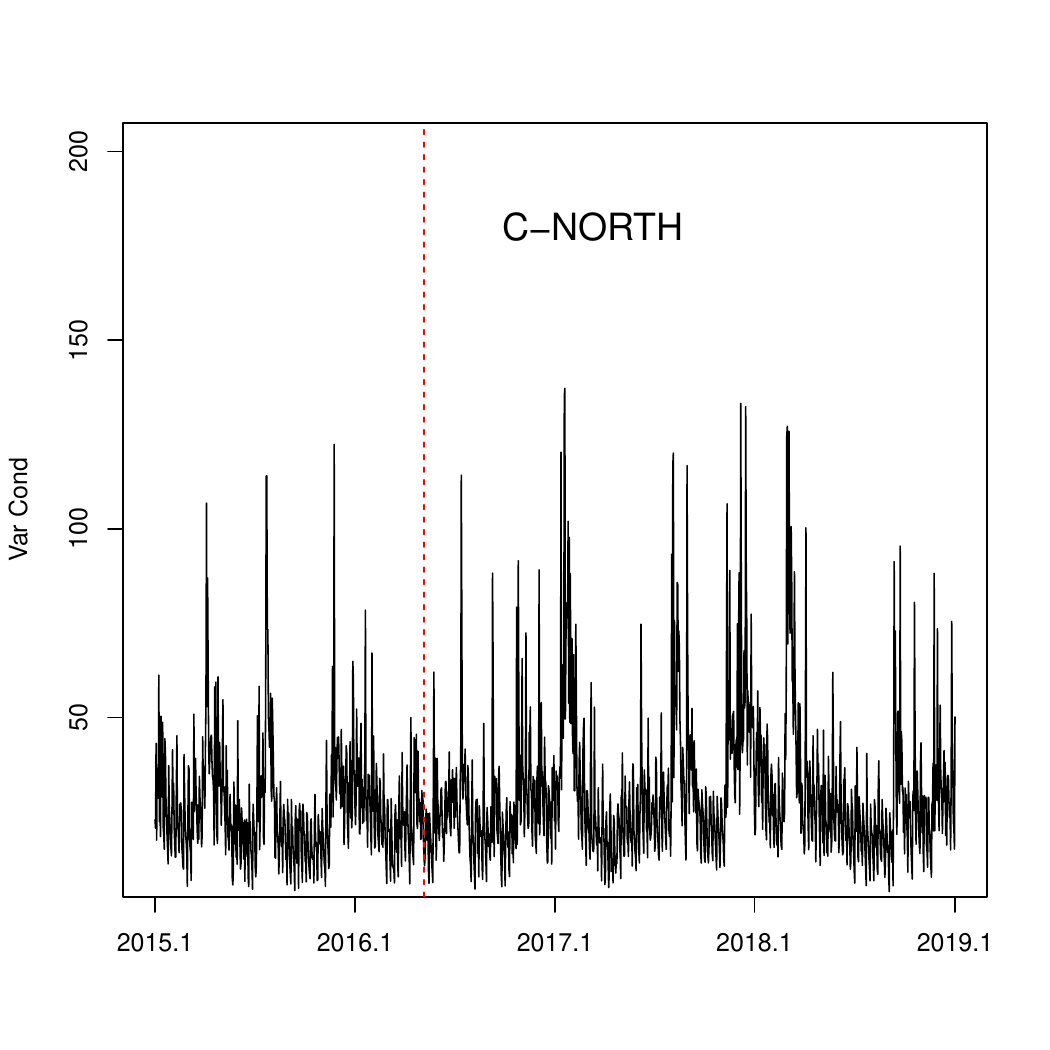}
	\includegraphics[width=6.6
	cm, height=5cm]{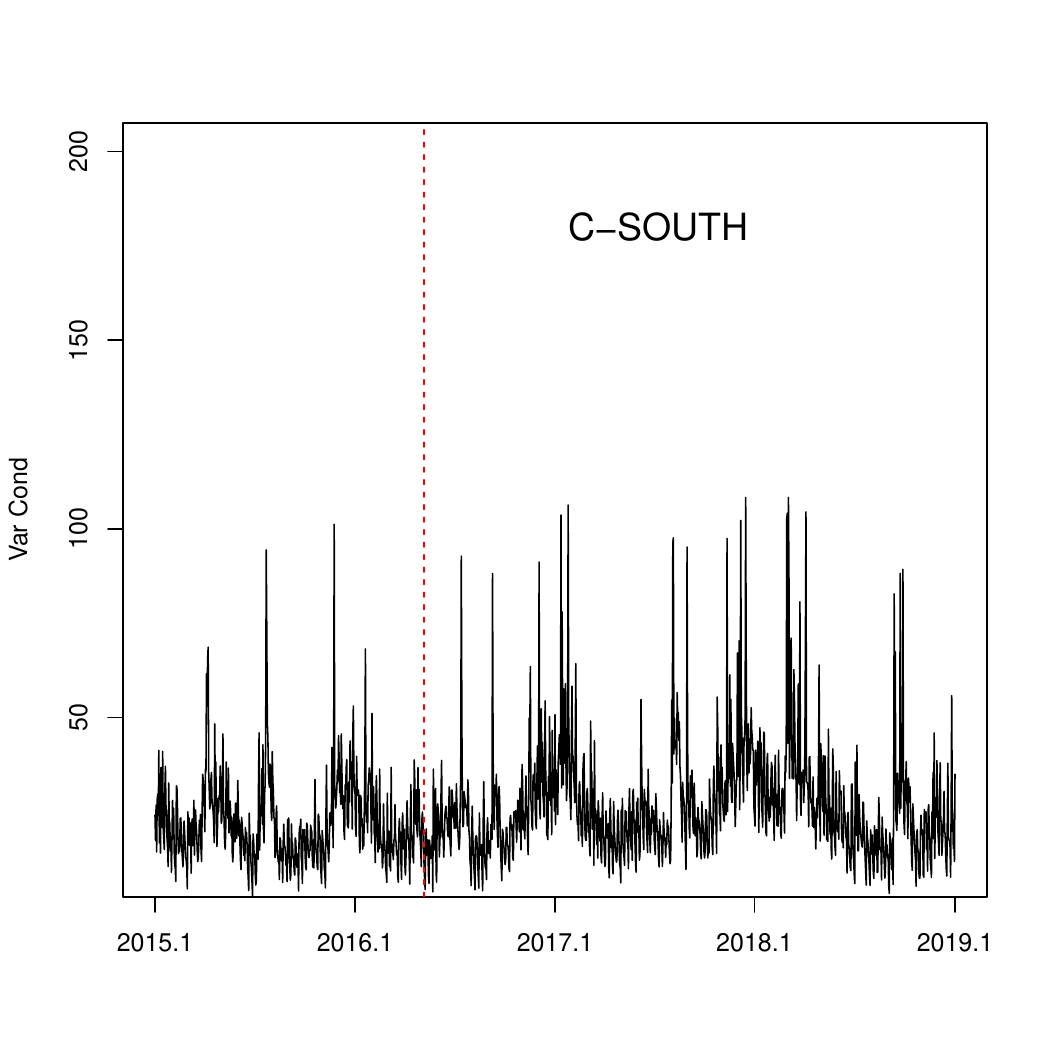}\\
	\includegraphics[width=6.6
	cm, height=5cm]{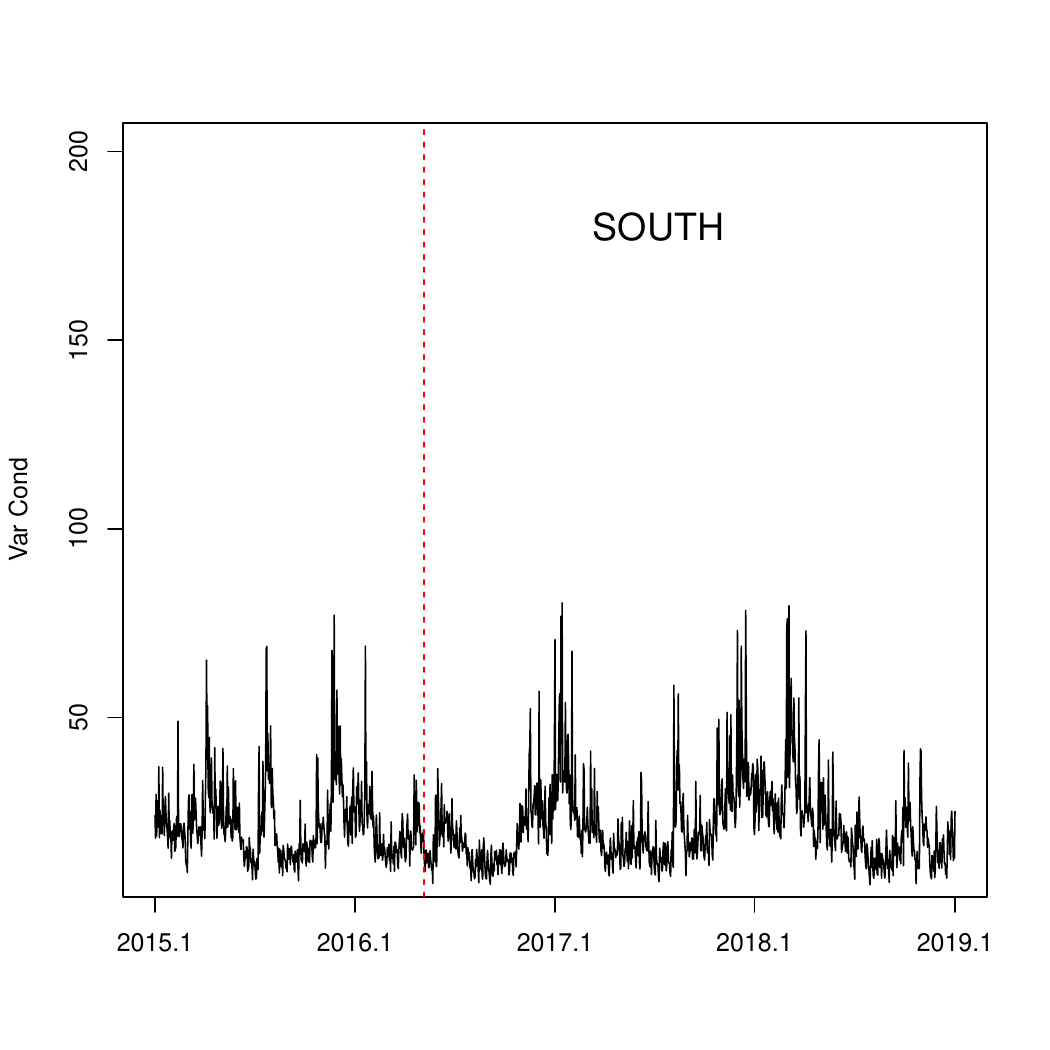}
	\includegraphics[width=6.6
	cm, height=5cm]{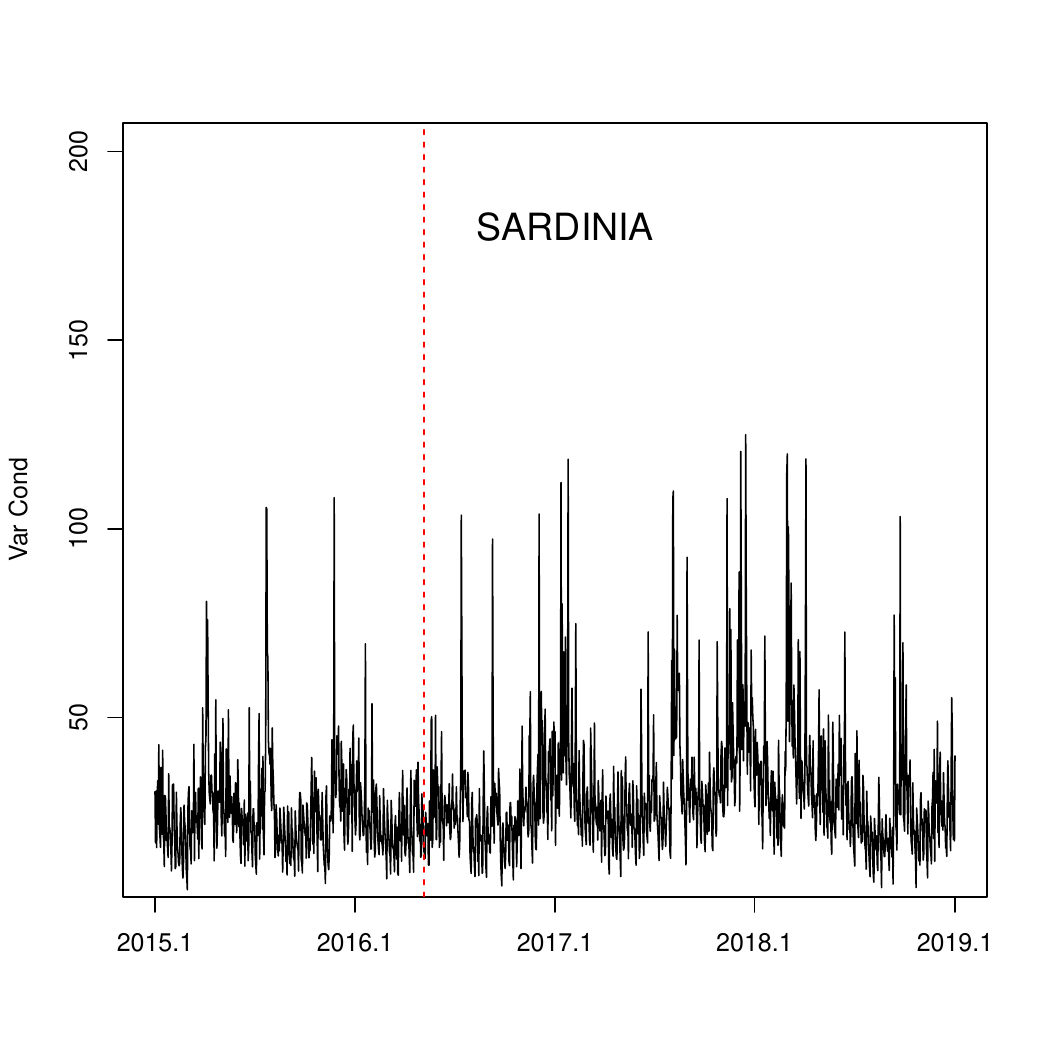}
	\caption{Estimated conditional variance for the six zones on the same scale.}
	\label{nonpar_varcond}
\end{figure}

The results are listed in Tab. \ref{anova}. For Sicily, the null hypothesis is rejected with a p-value equal to $0.008$, suggesting that the effect of the interconnector on price volatility is significant at the usual level of significance $5\%$. For all other zones, the test does not reject the null hypothesis at the same level, pushing us to conclude that the impacts represented by the estimated functions can be considered statistically null.\\
\begin{table}[H]
	\centering
	\begin{tabular}{lcccccc} \hline
		Zone & SICILY & NORTH & C-NORTH   & C-SOUTH & SOUTH & SARDINIA\\
		p-val 	& 0.008 & 0.324 &0.413& 0.250 & 0.261 & 0.160\\ 	\hline
	\end{tabular}
	\caption{ANOVA test for $H_0:$ No significant effect of the intervention variable.}
	\label{anova}
\end{table}
Figure \ref{nonpar_varcond} shows the estimated time series of the conditional variance for all zones.
Inherently with the ANOVA test and likewise with the semiparametric approach, only the volatility of Sicily has a clearly increasing level. In contrast, for the other zones, there are no clues of structural changes in the conditional variance time series.

\section{Conclusions}\label{sec:conclusions}
This paper analysed the impact of the introduction of the ``Sorgente-Rizziconi'' interconnector, linking Sicily with the Italian mainland, on price volatility in Sicily. \\
Volatility was modelled using a semi-parametric approach and a fully non-parametric approach for robustness checks. In the semi-parametric model, a sigmoidal-like impact on volatility was assumed. To ensure comparability, both models were also applied to other Italian market zones. \\  
The findings indicate that interconnector activation led to an increase in zonal price volatility in Sicily, whereas no significant changes were observed in the other zones.  
Moreover, following the introduction of the interconnector, the level-based volatility gradually increased until it reached a higher stable level. This behaviour is confirmed by the robustness check of the fully non-parametric approach. The magnitude of this change is estimated at approximately 10 euros by the semi-parametric model and approximately 17 euros by the fully non-parametric model. Aside from these numerical differences, due to the varying influence of volatility peaks in the parametric versus non-parametric modelling, the two approaches lead to the same qualitative conclusions. As expected, \textbf{H2} is verified in Sicily, while \textbf{H3} holds in the other market zones for the period 2015-2018. There is no evidence to support \textbf{H1}. As documented in the literature, the increase in price volatility can be attributed to the realignment of the Sicilian electricity market with that of the Italian mainland \citep{roques2021integration, CRETI20106966, NEWBERY2016253, CASSETTA2022112934}. This volatility can be interpreted as a signal of improved market competitiveness and of the improved ability of the transmission network to integrate local producers \citep{MALAGUZZIVALERI20094679,acemoglu2017competition,Bertolini20}.

Further support for this interpretation comes from the findings of \cite{Sapio2020} on the SAPEI interconnector in Sardinia, which show that volatility tends to be transmitted asymmetrically, from larger zones with a high share of intermittent renewable generation to smaller net-importing markets. This observed asymmetry underscores the structural dynamics of zonal electricity markets and highlights the role of interconnectors in redistributing volatility across regions.

However, the effects that interconnectors can have on prices and volatility are broad and context dependent, varying according to the specific characteristics of the markets involved. Although interconnectors often improve market integration and competition, their impact on price and volatility can differ significantly, depending on factors such as market structure, the share of renewables, and the direction of power flows \citep{CLEARY201638, Sapio16, PEAN2016307, RIES20161}.

This work can be extended in several directions. First, future research could adopt an hourly framework to investigate whether price volatility differs between peak and off-peak periods. Second, the role of the ``Sorgente-Rizziconi'' interconnector in driving volatility spillovers across market zones could be further examined, building on insights from \cite{Sapio2020}.

\paragraph{Declaration of interests} The authors report that financial support was provided by European Union - NextGenerationEU, Mission 4, Component 2, in the framework of the GRINS - Growing Resilient, INclusive and Sustainable project (GRINS PE00000018 - CUP C93C22005270001).

\paragraph{Acknowledgments} This study was funded by the European Union - NextGenerationEU, Mission 4, Component 2, in the framework of the GRINS - Growing Resilient, INclusive and Sustainable project (GRINS PE00000018 - CUP C93C22005270001). The views and opinions expressed are solely those of the authors and do not necessarily reflect those of the European Union, nor can the European Union be held responsible for them.\\
Marina Bertolini also acknowledges the support of the Levi Cases Centre of the University of Padova (project INCITE).

\paragraph{Data availability statement} Data on electricity prices are publicly available in \cite{mercatoelettrico2025}, while data on renewable energy (RES) production can be accessed from \cite{ENTSOE_homepage}.

\bibliographystyle{apalike}
\bibliography{cavo2_biblio}

\begin{thebibliography}{}

\bibitem[Acemoglu et~al., 2017]{acemoglu2017competition}
Acemoglu, D., Chernozhukov, V., Kasy, M., and Duflo, E. (2017).
\newblock Competition in electricity markets with renewable energy sources.
\newblock {\em The Energy Journal}, 38:137--155.
\newblock Accessed 5 June 2025.

\bibitem[Amado and Ter\"asvirta, 2013]{Amado_Terasvirta_2013}
Amado, C. and Ter\"asvirta, T. (2013).
\newblock Modelling volatility by variance decomposition.
\newblock {\em Journal of Econometrics}, 175:142--143.

\bibitem[Amado and Ter\"asvirta, 2017]{Amado_Terasvirta_2017}
Amado, C. and Ter\"asvirta, T. (2017).
\newblock Specification and testing of multiplicative time-varying garch models
  with applications.
\newblock {\em Econometric Reviews}, 36:421--446.

\bibitem[Bernardi and Lisi, 2020]{BERNARDI20}
Bernardi, M. and Lisi, F. (2020).
\newblock Point and interval forecasting of zonal electricity prices and demand
  using heteroscedastic models: The ipex case.
\newblock {\em Energies}, 13(23):6191.

\bibitem[Bertolini et~al., 2020]{Bertolini20}
Bertolini, M., Buso, M., and Greco, L. (2020).
\newblock Competition in smart distribution grids.
\newblock {\em Energy Policy}, 145:111729.

\bibitem[Cassetta et~al., 2022]{CASSETTA2022112934}
Cassetta, E., Nava, C.~R., and Zoia, M.~G. (2022).
\newblock Eu electricity market integration and cross-country convergence in
  residential and industrial end-user prices.
\newblock {\em Energy Policy}, 165:112934.

\bibitem[Cheng et~al., 2025]{CHENG2025110588}
Cheng, L., Peng, P., Lu, W., Sun, J., Wu, F., Shi, M., Yuan, X., and Chen, Y.
  (2025).
\newblock The evolutionary game equilibrium theory on power market bidding
  involving renewable energy companies.
\newblock {\em International Journal of Electrical Power \& Energy Systems},
  167:110588.

\bibitem[Cleary et~al., 2016]{CLEARY201638}
Cleary, B., Duffy, A., Bach, B., Vitina, A., O’Connor, A., and Conlon, M.
  (2016).
\newblock Estimating the electricity prices, generation costs and co2 emissions
  of large scale wind energy exports from ireland to great britain.
\newblock {\em Energy Policy}, 91:38--48.

\bibitem[Codegoni, 2016]{Codegoni2016}
Codegoni, A. (2016).
\newblock {Prezzi elettrici e caso Sicilia: ma il cavo Sorgente-Rizziconi
  funziona o no?}
\newblock Consultato il 2 giugno 2025.

\bibitem[Creti et~al., 2010]{CRETI20106966}
Creti, A., Fumagalli, E., and Fumagalli, E. (2010).
\newblock Integration of electricity markets in europe: Relevant issues for
  italy.
\newblock {\em Energy Policy}, 38(11):6966--6976.
\newblock Energy Efficiency Policies and Strategies with regular papers.

\bibitem[Denny et~al., 2010]{DENNY20106946}
Denny, E., Tuohy, A., Meibom, P., Keane, A., Flynn, D., Mullane, A., and
  O’Malley, M. (2010).
\newblock The impact of increased interconnection on electricity systems with
  large penetrations of wind generation: A case study of ireland and great
  britain.
\newblock {\em Energy Policy}, 38(11):6946--6954.
\newblock Energy Efficiency Policies and Strategies with regular papers.

\bibitem[Dudjak et~al., 2021]{DUDJAK2021117434}
Dudjak, V., Neves, D., Alskaif, T., Khadem, S., Pena-Bello, A., Saggese, P.,
  Bowler, B., Andoni, M., Bertolini, M., Zhou, Y., Lormeteau, B., Mustafa,
  M.~A., Wang, Y., Francis, C., Zobiri, F., Parra, D., and Papaemmanouil, A.
  (2021).
\newblock Impact of local energy markets integration in power systems layer: A
  comprehensive review.
\newblock {\em Applied Energy}, 301:117434.

\bibitem[{ENTSO‑E}, 2025]{ENTSOE_homepage}
{ENTSO‑E} (2025).
\newblock Entso‑e – building a clean and competitive power system for
  europe.
\newblock \url{https://www.entsoe.eu/}.
\newblock Accessed: 2025‑07‑31.

\bibitem[{European Commission}, 2014]{EC2014climateframework}
{European Commission} (2014).
\newblock A policy framework for climate and energy in the period from 2020 to
  2030.
\newblock
  \url{https://eur-lex.europa.eu/legal-content/EN/ALL/?uri=CELEX:52014DC0015}.

\bibitem[{European Commission}, 2023]{eu_grid_action_plan_2023}
{European Commission} (2023).
\newblock Commission communication: Grids, the missing link – an eu action
  plan for grids.
\newblock
  \url{https://eur-lex.europa.eu/legal-content/EN/TXT/?uri=COM:2023:757:FIN}.

\bibitem[{European Commission}, 2025]{EU_ElectricityMarketDesign}
{European Commission} (2025).
\newblock Electricity market design.
\newblock
  \url{https://energy.ec.europa.eu/topics/markets-and-consumers/electricity-market-design_en}.
\newblock Accessed: 27 July 2025.

\bibitem[{European Subsea Cables Association}, 2025]{ESCA_SubmarineCables}
{European Subsea Cables Association} (2025).
\newblock Submarine power cables.
\newblock \url{https://www.escaeu.org/articles/submarine-power-cables/}.
\newblock Accessed: 2025-10-14.

\bibitem[{European Union Parliament}, 2018]{EU2018Reg1999_governance}
{European Union Parliament} (2018).
\newblock Regulation (eu) 2018/1999 of the european parliament and of the
  council of 11 december 2018 on the governance of the energy union and climate
  action.
\newblock
  \url{https://eur-lex.europa.eu/legal-content/EN/TXT/PDF/?uri=CELEX:32018R1999}.

\bibitem[{European Union Parliament}, 2024]{EU2024directive1711}
{European Union Parliament} (2024).
\newblock Directive (eu) 2024/1711 of the european parliament and of the
  council of 13 june 2024.
\newblock Official Journal of the European Union.

\bibitem[Fabra, 2023]{FABRA2023}
Fabra, N. (2023).
\newblock Reforming european electricity markets: Lessons from the energy
  crisis.
\newblock {\em Energy Economics}, 126:106963.

\bibitem[Fischhendler et~al., 2015]{Fischhendler2015}
Fischhendler, I., Nathan, D., and Boymel, D. (2015).
\newblock Marketing renewable energy through geopolitics: Solar farms in
  israel.
\newblock {\em Global Environmental Politics}, 15(2):98--120.

\bibitem[{GME}, 2025]{mercatoelettrico2025}
{GME} (2025).
\newblock Zonal electricity prices — market results.
\newblock
  \url{https://www.mercatoelettrico.org/en-us/Home/Results/Electricity/MGP/Results/ZonalPrices}.
\newblock Accessed: 2025-07-29.

\bibitem[Hastie et~al., 2001]{Hastie_etal_2001}
Hastie, T., Tibshirani, R., and Friedman, J. (2001).
\newblock {\em The Elements of Statistical Learning: Data Mining, Inference,
  and Prediction}.
\newblock Springer.

\bibitem[Jansen et~al., 2022]{JANSEN2022}
Jansen, M., Duffy, C., Green, T.~C., and Staffell, I. (2022).
\newblock Island in the sea: The prospects and impacts of an offshore wind
  power hub in the north sea.
\newblock {\em Advances in Applied Energy}, 6:100090.

\bibitem[Li et~al., 2025]{LI2025101808}
Li, J., Li, G., Song, F., and Feng, Y. (2025).
\newblock Addressing the energy trilemma: Progress and evaluation of
  electricity market reform in china.
\newblock {\em Energy for Sustainable Development}, 88:101808.

\bibitem[Lisi and Pelagatti, 2018]{LISI18}
Lisi, F. and Pelagatti, M.~M. (2018).
\newblock Component estimation for electricity market data: Deterministic or
  stochastic?
\newblock {\em Energy Economics}, 74:13--37.

\bibitem[{Lo Prete} et~al., 2025]{LOPRETE2025108640}
{Lo Prete}, C., Palmer, K., and Robertson, M. (2025).
\newblock Time for a market upgrade? a review of wholesale electricity market
  designs for the future.
\newblock {\em Energy Economics}, 148:108640.

\bibitem[Lobato et~al., 2017]{LOBATO2017192}
Lobato, E., Sigrist, L., and Rouco, L. (2017).
\newblock Value of electric interconnection links in remote island power
  systems: The spanish canary and balearic archipelago cases.
\newblock {\em International Journal of Electrical Power \& Energy Systems},
  91:192--200.

\bibitem[Álvaro Cartea and González-Pedraz, 2012]{CARTEA201214}
Álvaro Cartea and González-Pedraz, C. (2012).
\newblock How much should we pay for interconnecting electricity markets? a
  real options approach.
\newblock {\em Energy Economics}, 34(1):14--30.

\bibitem[{Malaguzzi Valeri}, 2009]{MALAGUZZIVALERI20094679}
{Malaguzzi Valeri}, L. (2009).
\newblock Welfare and competition effects of electricity interconnection
  between ireland and great britain.
\newblock {\em Energy Policy}, 37(11):4679--4688.

\bibitem[Meneguzzo et~al., 2016]{Meneguzzo2016}
Meneguzzo, F., Ciriminna, R., Albanese, L., and Pagliaro, M. (2016).
\newblock The remarkable impact of renewable energy generation in sicily onto
  electricity price formation in italy.
\newblock {\em Energy Science \& Engineering}, 4(3):194--204.

\bibitem[Milstein and Tishler, 2015]{MILSTEIN201570}
Milstein, I. and Tishler, A. (2015).
\newblock Can price volatility enhance market power? the case of renewable
  technologies in competitive electricity markets.
\newblock {\em Resource and Energy Economics}, 41:70--90.

\bibitem[Mu\u{g}alo\u{g}lu et~al., 2026]{MUGALOGLU2026}
Mu\u{g}alo\u{g}lu, E., K{\i}l{\i}\c{c}, E., Keskin, H., and Sel\c{c}uklu, S.~B.
  (2026).
\newblock Dynamics of electricity price volatility and its impacts on energy
  investments.
\newblock {\em Renewable Energy}, 256:124383.

\bibitem[Newbery et~al., 2016]{NEWBERY2016253}
Newbery, D., Strbac, G., and Viehoff, I. (2016).
\newblock The benefits of integrating european electricity markets.
\newblock {\em Energy Policy}, 94:253--263.

\bibitem[Pean et~al., 2016]{PEAN2016307}
Pean, E., Pirouti, M., and Qadrdan, M. (2016).
\newblock Role of the gb-france electricity interconnectors in integration of
  variable renewable generation.
\newblock {\em Renewable Energy}, 99:307--314.

\bibitem[Pinilla and Negr\'in, 2021]{Pinilla2021}
Pinilla, J. and Negr\'in, M. (2021).
\newblock Non-parametric generalized additive models as a tool for evaluating
  policy interventions.
\newblock {\em Mathematics}, 9:299.

\bibitem[Proedrou, 2012]{Proedrou201215}
Proedrou, F. (2012).
\newblock Re-conceptualising the energy and security complex in the eastern
  mediterranean.
\newblock {\em Cyprus Review}, 24(2):15 – 28.
\newblock Cited by: 8.

\bibitem[Rettig et~al., 2023]{RETTIG2023113732}
Rettig, E., Fischhendler, I., and Schlecht, F. (2023).
\newblock The meaning of energy islands: Towards a theoretical framework.
\newblock {\em Renewable and Sustainable Energy Reviews}, 187:113732.

\bibitem[Ries et~al., 2016]{RIES20161}
Ries, J., Gaudard, L., and Romerio, F. (2016).
\newblock Interconnecting an isolated electricity system to the european
  market: The case of malta.
\newblock {\em Utilities Policy}, 40:1--14.

\bibitem[Roques, 2021]{roques2021integration}
Roques, F. (2021).
\newblock The integration of european electricity markets--achievements to date
  and way forward.
\newblock In {\em Energy, COVID, and Climate Change, 1st IAEE Online
  Conference, June 7-9, 2021}. International Association for Energy Economics.

\bibitem[Sapio and Spagnolo, 2016]{Sapio16}
Sapio, A. and Spagnolo, N. (2016).
\newblock Price regimes in an energy island: Tacit collusion vs. cost and
  network explanations.
\newblock {\em Energy Economics}, 55:157--172.

\bibitem[Sapio and Spagnolo, 2020]{Sapio2020}
Sapio, A. and Spagnolo, N. (2020).
\newblock The effect of a new power cable on energy prices volatility
  spillovers.
\newblock {\em Energy Policy}, 144:111488.

\bibitem[Shortall and Kharrazi, 2017]{SHORTALL2017101}
Shortall, R. and Kharrazi, A. (2017).
\newblock Cultural factors of sustainable energy development: A case study of
  geothermal energy in iceland and japan.
\newblock {\em Renewable and Sustainable Energy Reviews}, 79:101--109.

\bibitem[{Terna S.p.A.}, 2016]{Terna2016}
{Terna S.p.A.} (2016).
\newblock {L'energia che ci unisce: il "ponte dell'energia" tra Sicilia e
  Calabria}.
\newblock Technical report, Terna S.p.A.
\newblock Consultato il 2 giugno 2025.

\bibitem[Tishler et~al., 2008]{TISHLER20081625}
Tishler, A., Milstein, I., and Woo, C.-K. (2008).
\newblock Capacity commitment and price volatility in a competitive electricity
  market.
\newblock {\em Energy Economics}, 30(4):1625--1647.

\bibitem[Trovato, 2016]{Trovato2016}
Trovato, M. (2016).
\newblock {Ma l’elettrodotto Sorgente-Rizziconi fa risparmiare o no?}
\newblock Consultato il 2 giugno 2025.

\bibitem[Tsaousoglou et~al., 2022]{TSAOUSOGLOU2022}
Tsaousoglou, G., Giraldo, J.~S., and Paterakis, N.~G. (2022).
\newblock Market mechanisms for local electricity markets: A review of models,
  solution concepts and algorithmic techniques.
\newblock {\em Renewable and Sustainable Energy Reviews}, 156:111890.

\bibitem[Turvey, 2006]{TURVEY20061457}
Turvey, R. (2006).
\newblock Interconnector economics.
\newblock {\em Energy Policy}, 34(13):1457--1472.

\bibitem[Wang et~al., 2024]{WANG2024144343}
Wang, G., Sbai, E., Wen, L., and {Selena Sheng}, M. (2024).
\newblock The impact of renewable energy on extreme volatility in wholesale
  electricity prices: Evidence from organisation for economic co-operation and
  development countries.
\newblock {\em Journal of Cleaner Production}, 484:144343.

\bibitem[Warneryd and Karltorp, 2022]{WARNERYD2022}
Warneryd, M. and Karltorp, K. (2022).
\newblock Microgrid communities: disclosing the path to future system-active
  communities.
\newblock {\em Sustainable Futures}, 4:100079.

\bibitem[Weinhold and Mieth, 2021]{Weinhold21}
Weinhold, R. and Mieth, R. (2021).
\newblock Power market tool (pomato) for the analysis of zonal electricity
  markets.
\newblock {\em SoftwareX}, 16:100870.

\bibitem[Yang et~al., 2024]{YANG2024}
Yang, Y., Xia, S., Huang, P., and Qian, J. (2024).
\newblock Energy transition: Connotations, mechanisms and effects.
\newblock {\em Energy Strategy Reviews}, 52:101320.

\end{thebibliography}

\end{document}